\documentclass[12pt]{article}

\usepackage{PRIMEarxiv}

\usepackage[utf8]{inputenc} 
\usepackage[T1]{fontenc} 
\usepackage{amsthm,amsmath}
\usepackage{amsthm}
\usepackage{natbib}
\usepackage{hyperref}       
\usepackage{url}            
\usepackage{booktabs}       
\usepackage{amsfonts}       
\usepackage{nicefrac}       
\usepackage{microtype}      
\usepackage{fancyhdr}       
\usepackage{graphicx}
\usepackage[affil-it]{authblk}
\usepackage{algorithmic}
\usepackage[ruled,vlined]{algorithm2e}
\usepackage{caption}
\usepackage{subcaption}
\graphicspath{{media/}}     

\DeclareMathOperator*{\argmin}{arg\,min}
\DeclareMathOperator*{\KL}{KL}

\newcommand{\CytOpT}{\texttt{CytOpT}}
\newcommand{\OptimalFlow}{\texttt{OptimalFlow}}

\theoremstyle{rmq}

\title{CytOpT: Optimal Transport with Domain Adaptation for Interpreting Flow Cytometry data}

\author[*,1,2]{\large Paul Freulon}
\author[1,2]{{\large J\'er\'emie Bigot}}
\author[1,3,4]{\large Boris P.\ Hejblum}

\affil[1]{Universit\'{e} de Bordeaux, Bordeaux, 33000, France.}
\affil[2]{Institut de Math\'ematiques de Bordeaux et CNRS  (UMR 5251), 33400 Talence, France.}
\affil[3]{Bordeaux Population Health Research Center Inserm U1219, Inria SISTM, 33000 Bordeaux, France.}
\affil[4]{Vaccine Research Institute (VRI), 94010 Cr\'eteil, France.}
\affil[*]{corresponding author: paul.freulon@math.u-bordeaux.fr}

\begin{document}
\maketitle

\begin{abstract}

The automated analysis of flow cytometry measurements is an active research field. We introduce a new algorithm, referred to as \texttt{CytOpT}, using regularized optimal transport to directly estimate the different cell population proportions from a biological sample characterized with flow cytometry measurements. We rely on the regularized Wasserstein metric to compare cytometry measurements from different samples, thus accounting for possible mis-alignment of a given cell population across samples (due to technical variability from the technology of measurements). In this work, we rely on a supervised learning technique based on the Wasserstein metric that is used to estimate an optimal re-weighting of class proportions in a mixture model from a source distribution (with known segmentation into cell sub-populations) to fit a target distribution with unknown segmentation. Due to the high-dimensionality of flow cytometry data, we use stochastic algorithms to approximate the regularized Wasserstein metric to solve the optimization problem involved in the estimation of optimal weights representing the cell population proportions in the target distribution. Several flow cytometry data sets are used to illustrate the performances of \texttt{CytOpT} that are also compared  to those of existing algorithms for automatic gating based on supervised learning.
\end{abstract}


\section{Introduction}

\subsection{Flow Cytometry Data Analysis}
Flow cytometry is a high-throughput biotechnology used to characterize a large amount of cells from a biological sample. Flow cytometry is paramount to many biological and immunological research with applications, for instance, in the monitoring of the immune system of HIV patients by counting the number of CD4 cells.

The first step to characterize cells from a biological sample with flow cytometry is to stain those cells. 
Specifically, cells are stained with multiple fluorescently-conjugated monoclonal antibodies directed to the cellular markers of interest. Then, the cells flow one by one through the cytometer laser beam. The scattered light is characteristic to the biological markers of the cells \citep{aghaeepour2013critical}. Thus, from a biological sample analyzed by a flow cytometer, we get a data set $X_1,...,X_I$ where each observation $X_i$ corresponds to a single cell crossing the laser beam. For an observation $X_i \in \mathbb{R}^d$, the coordinate $X_i^{(m)}$ corresponds to the light intensity emitted by the fluorescent antibody attached to the biological marker $m$. Interestingly, such a data set may also be considered as a discrete probability distribution $\frac{1}{I} \sum_{i=1}^{I} \delta_{X_i}$ with support in $\mathbb{R}^d$, this is the point of view taken in this paper.

With flow cytometry one can assess individual characteristics at a cellular level for samples composed of hundreds of thousands of cells. The development of this technology now leads to state-of-the-art cytometers that can measure up to 50 biological markers at once on a single cell \citep{saeys2016computational}.
The constant increase of the number of measured markers allows to identify more precisely the different cell populations and subtypes present in the biological sample.

The analysis of cytometry data is generally done manually, by drawing
geometric shapes (referred to as ``gates'') around populations of interest in a sequence of two-dimensional data projections. Such manual analysis features several drawbacks: i) it is extremely time-consuming ; ii) manual gating lacks reproducibility across different operators \citep{aghaeepour2013critical}. To overcome these shortcomings, several automated methods have been proposed \citep{aghaeepour2013critical}. Those automated approaches aim at a clustering of the flow cytometry data to derive the proportions of the cell populations that are in the biological sample. Some methods follow an unsupervised approach. For instance, \texttt{FlowMeans} \citep{aghaeepour2011rapid} is an automated method based on the K-means designed for flow cytometry data. We also mention {\tt Cytometree} \citep{Cytometree}, an algorithm based on the construction of a binary tree. Methods which perform model based clustering have also been proposed, see e.g.\ \citet{Flow_Peak,NP_flow}.
To improve the accuracy of the classification, supervised machine learning techniques have been applied to flow cytometry data analysis. Among those techniques, one may cite {\tt DeepCyTOF} introduced in \citet{deep_cytof}  which is based on deep-learning algorithms to gate cytometry data. In \citet{flow_learn}, the authors introduced {\tt flowlearn}, a method that uses manually gated samples to predict gates on other samples.  We also mention a new supervised approach named {\tt OptimalFlow} developed by \citet{Optimalflow} that relies on the Wasserstein distance to quantify discrepancies between cytometry data sets.

In spite of numerous efforts to automate cytometry data analysis, manual gating remains the gold-standard for benchmarking. the fact that automated methods have not outdone manual gating can be explained, at least in part, by the significant variability of flow cytometry data.
This variability is two fold; first, it is induced by biological heterogeneity across the samples analyzed \citep{hahne2010per}. For instance, it is likely that subtypes proportions within the biological sample of a healthy patient will differ from the subtypes proportions within the biological sample of a sick patient. In addition to this variability of the biological phenomena of interest, technical variability appears during the process of flow cytometry analysis. For instance, differences in the staining procedure, in the data acquisition settings or cytometers performances are very likely to happen and to lead to undesirable variability between flow cytometry data \citep{maecker2010model}. 

\subsection{Optimal transport in Statistics}

To tackle the matter of Flow Cytometry data analysis, this work proposes to make use of tools that come from the theory of optimal transport. Optimal transport has recently gained interest in machine learning and statistics. Indeed, the introduction of approximate solvers that for large dimension problems allowed to move beyond the high computational cost of optimal transport. Thus, optimal transport has found various applications in machine learning for regression \citep{Wasserstein_regression}, classification \citep{Wasserstein_discriminant_analysis} and generative modeling \citep{Wasserstein_gan}. Those computational progresses in optimal transport have also allowed its use in imaging sciences \citep{Convolutional_wasserstein}.
Recent progresses in applied optimal transport has been fueled by the development of efficient, large-scale optimization algorithms for problems in this domain. In particular, our method rely on stochastic algorithms \citep{Stochastic_Genevay,Stochastic_Bigot_Bercu} to 	alleviate the computational cost of optimal transport. 
In relationship with our field of research, and to make the point that optimal transport tools are now applied on real-life data, we mention two single cell algorithms based on optimal transport lately published. First \citet{schiebinger2019optimal} model the development of cells in order to understand their differentiation process. And in \citet{li2019optimal}, the authors propose a method that deals with the issue of cluster alignment with the purpose to discriminate between intrinsic clusters and spurious clusters.

\subsection{An illustrative data set}\label{subsec:data_set}

\begin{figure}[!ht]
	\begin{subfigure}{.5\textwidth}
		\hspace{-2cm}
		\includegraphics[width=1.3\linewidth]{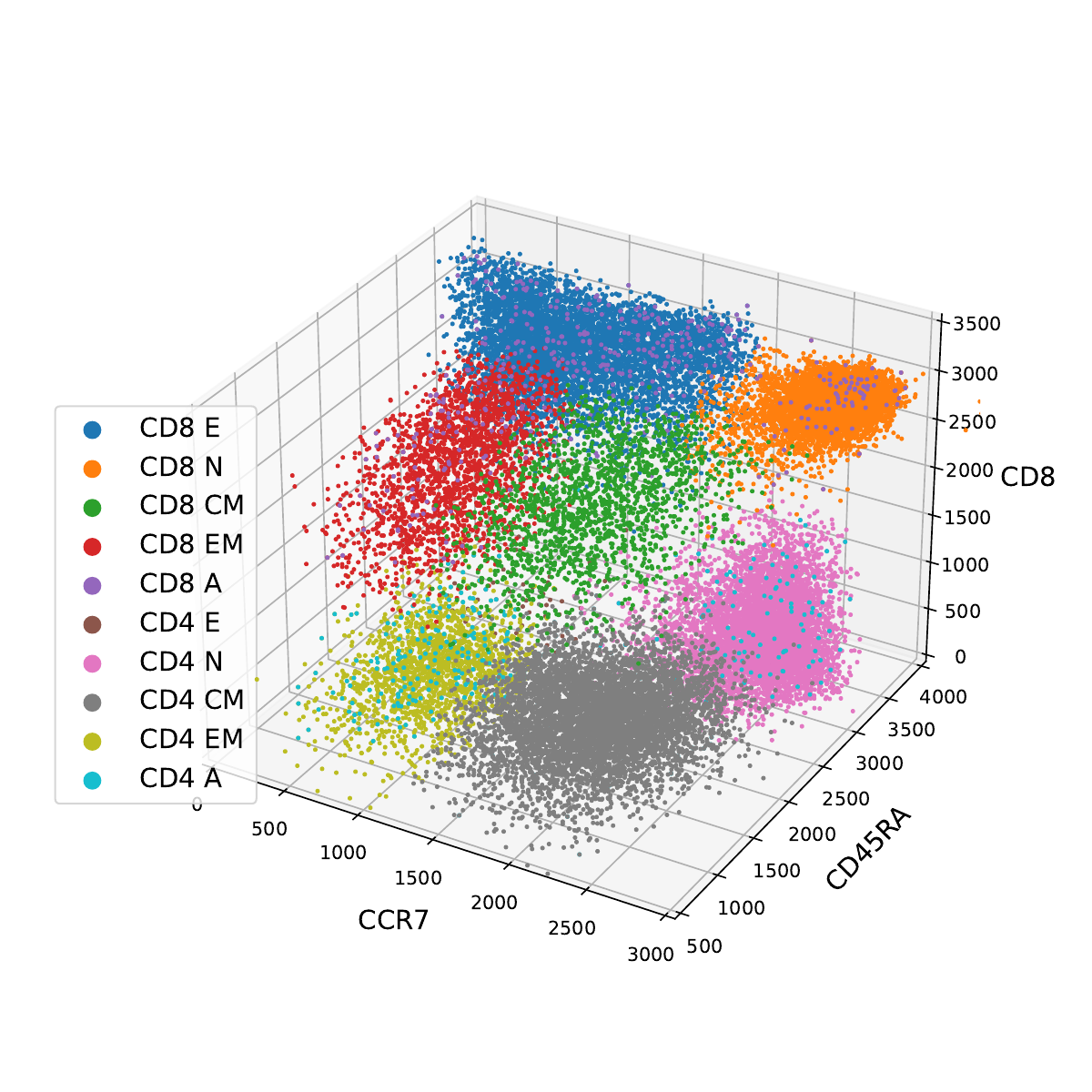}
	\end{subfigure}%
	\begin{subfigure}{.5\textwidth}
		\hspace*{0.5cm}
		\includegraphics[width=1\linewidth]{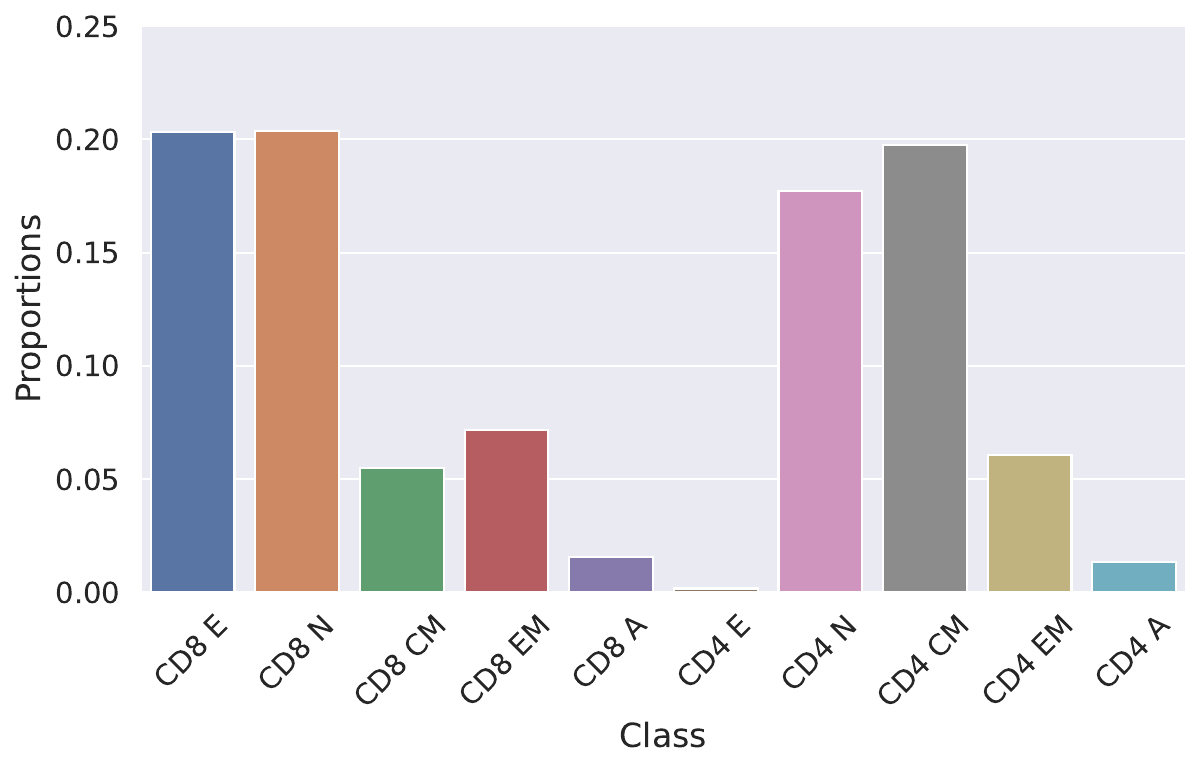}
	\end{subfigure}
	\vspace{-1.2cm}
	\caption{\footnotesize\textbf{Example flow cytometry data set of a biological sample from Stanford patient 1 replicate A.} Left: Manual clustering of the cytometry data. Existing automated methods target a clustering in order to derive the class proportions. Right: Class proportion derived from the manual gating. Our method {\tt CytOpT} aims at going straight to the estimation of class proportions without clustering the cytometry data.}
	\label{fig:Stanford1A}
\end{figure}

As an illustrative example, we shall analyze in this paper the flow cytometry data from the T-cell panel of the Human Immunology Project Consortium (HIPC) \--- publicly available on ImmuneSpace \citep{Data_HIPC}. Seven laboratories stained three replicates (denoted A, B, and C) of three cryo-preserved biological samples denoted patient 1, 2, and 3 (e.g.\ cytometry measurements from the Stanford laboratory for replicate C from patient 1 will be denoted as the ``Stanford1C'' data set). After performing  cytometry measurements in each center, the resulting FCS files were manually gated centrally for quantifying 10 cell populations: CD4 Effector (CD4 E), CD4 Naive (CD4 N), CD4 Central memory (CD4 CM), CD4 Effector memory (CD4 EM), CD4 Activated (CD4 A), CD8 Effector (CD8 E), CD8 Naive (CD8 N), CD8 Central memory (CD8 CM), CD8 Effector memory (CD8 EM) and CD8 Activated (CD8 A). Hence, for these data sets, a manual clustering is at our disposal to evaluate the performances of automatic gating methods. The flow cytometry data sets built from these 10 sub populations have a size that range from  15 554 observations for the smallest data set to 112 318 observations for the largest data set. For each cell, seven biological markers have been measured; it leads to cytometry observations $X_i$ that belong to $\mathbb{R}^d$ with $d=7$. A 3D projection using three markers  is displayed in Figure \ref{fig:Stanford1A} for the ``Stanford1A'' data set with the corresponding manual gating into 10 clusters.
Additionally, we also benchmarked our method on the flow cytometry data sets used in \cite{Optimalflow}.

\subsection{Main contributions}

We propose a new supervised method to estimate the relative proportions of the different cell sub-types in a biological sample analyzed by a cytometer. Our approach aims at finding an optimal re-weighting of class proportions in a mixture model between a source data set (with known segmentation into cell sub-populations) to fit a target data set with unknown segmentation. This estimation of the class proportions is done without any preliminary clustering of the target cytometry data. To the best of our knowledge, all the previous automated methods are based on the classification of the cells of the biological sample to deduce a class proportions estimation. However, from a clinical perspective, the relevant information is the class proportions \citep{maecker2010model}, while the clustering of individual cells is simply a means to an end to get there. Our method can nonetheless be extended in order to obtain a clustering of cytometry data.

We believe that going straight to the estimation of class proportions is an original approach to tackle the issue of flow cytometry data analysis. Hence, the result of our algorithm is not a clustering of the cells analyzed by the cytometer but a vector $\hat{\pi}=(\hat{\pi}_1,...,\hat{\pi}_K)$, where each coefficient $\hat{\pi}_k$ amounts to estimation of the percentage of the $k^{\text{th}}$ population cell among all the cells analyzed in the target sample. To do so, we rely on tools derived from optimal transport and regularized Wasserstein distance between probability measures.

In this work, we propose to estimate the proportions $\pi=(\pi_1,...,\pi_K)$ of sub-populations cell in a target data set (whose gating is unknown) by minimizing  the regularized Wasserstein distance between the target distribution, and a re-weighted source distribution (from a source data set whose gating into $K$ sub-populations cell of interest is known). Since the Wasserstein distance is able to capture the underlying geometry of the measurements, it has the ability to make meaningful comparisons between distributions whose supports do not overlap and to quantify potential spatial shifts. Our work demonstrates the benefits of using Wasserstein distance for the analysis of the  cytometry data as it allows to handle the technical variability induced by different settings of measurements.

\subsection{Organization of the paper}

In Section \ref{sec:math_framework} we present the regularized Wasserstein distance and the stochastic procedure that we leverage to compute it. Section \ref{sec:method} details our estimation method named \texttt{CytOpT} that yields an estimate of the class proportions in an unsegmented data set. Section \ref{sec:simulation_study} is devoted to a thorough simulation study of \texttt{CytOpT} where it is compared to classic classification methods and its robustness is assessed. Finally, in Section \ref{sec:results} we demonstrate the performance of \CytOpT{} on the T-cell panel proposed by the Human Immunology Project Consortium (HIPC) described previously, and we compare the performance of \CytOpT, with existing methods.

\section{Mathematical framework}\label{sec:math_framework}

Optimal transport allows the definition of a metric between two probability distributions $\alpha$ and $\beta$ supported on $\mathbb{R}^d$. This metric is informally defined as the lowest cost to move the mass from one probability measure, the source measure $\alpha$, onto the other, the target measure $\beta$. As optimal transport handles probability distributions, we must describe a cytometry data set $X_1,...,X_I$ where each $X_i$ belongs to $\mathbb{R}^d$ as a probability distribution in $\mathbb{R}^d$. A natural choice to move from observations to probability measure
is to use the empirical measure. Therefore, for a data set $X_1,...,X_I$, we consider its empirical measure defined as 
$\frac{1}{I} \sum_{i=1}^I \delta_{X_i}.$

\subsection{Wasserstein distance and its regularized counterpart}\label{subsec:wassertein_metrics}

In this work, we shall consider optimal transport between discrete measures on $\mathbb{R}^d$, and we denote by $\alpha = \sum_{i=1}^I a_i\delta_{x_i}$ and $\beta=\sum_{j=1}^J b_j\delta_{y_j}$ two such measures. Note that $a_i$, the probability associated to the point $x_i$, is not necessarily equal to $1/I$ as a re-weighting of the source measure will be considered. Optimal transport then seeks a plan that moves the probability measure $\alpha$ to the probability measure $\beta$ with a minimum cost. The transportation cost is encoded by a function $c: \mathbb{R}^d \times \mathbb{R}^d \rightarrow \mathbb{R}_+$ where $c(x,y)$ represents the cost to move one unit of mass from $x$ to $y$. In the discrete setting the transportation cost is encoded by a matrix $C \in \mathbb{R}^{I \times J}$ where $C_{i,j} = c(x_i,y_j)$. Once set, the cost matrix $C$ ensures that the transportation is carried out between a source distribution with support $x_1,...,x_I$ and a target distribution with support $y_1,...,y_J$. Then, to ensure the conservation of mass, i.e. the transport of a distribution with weights $a_1,...,a_I$ toward a distribution with weights $b_1,...,b_J$, we introduce the set $\Pi(a, b) = \{ P \in \mathbb{R}^{I\times J}: P1_J = a ~\text{and}~ P^T 1_I = b  \}$ of all the coupling matrices between $a$ and $b$. Thus, minimizing the transportation cost from $\alpha$ to $\beta$ boils down to the following optimization problem:
\begin{equation}\label{eq:kantorovich_problem}
	W_c(\alpha, \beta) = \inf_{\pi \in \Pi(a, b)} \sum_{i=1}^I \sum_{j=1}^J C_{i,j} P_{i,j} = \inf_{\pi \in \Pi(a, b)} \langle C, P \rangle
\end{equation}
All along this work the coefficient $C_{i,j}$ of the cost matrix will be defined as the squared Euclidean distance between $x_i$ and $y_j$, i.e. $C_{i,j} = ||x_i - y_j||_2^2$, and we use the notation $W(\alpha, \beta) = W_2^2(\alpha, \beta)$. In spite of appealing theoretical properties (see e.g. \citet{santambrogio2015optimal}) Wasserstein metrics drag a computational burden. Indeed, the cost to evaluate the Wasserstein distance between two discrete probability distributions with support of equal size $N$ is generally of order $O(N^3\log(N))$ \citep{Lightspeed}. To allow the evaluation of the Wasserstein distance at a lower cost, \citet{Lightspeed} has proposed to add an entropic regularization term to the linear optimization problem \eqref{eq:kantorovich_problem} in order to reach an approximate solution with faster computations. This regularized optimal transport problem reads:
\begin{equation}\label{eq:regularized_problem}
	W^{\varepsilon}(\alpha, \beta) = \inf_{\pi \in \Pi(a, b)} \langle C, P \rangle + \varepsilon H(P),
\end{equation}
where the entropy $H: \mathbb{R}_+^{I\times J} \rightarrow \mathbb{R}$ is defined for $P \in \mathbb{R}_+^{I \times J}$ by $H(P) = \sum_{i,j}\left(\log\left(P_{i,j}\right)\right.$ $\left. - 1\right)P_{i,j}$, and $\varepsilon > 0$ is the regularization parameter. The entropic regularization of the Kantorovich problem \eqref{eq:regularized_problem} leads to an approximation of the Wasserstein distance which can be calculated in $O(N^2\log(N))$ operations if the two distributions have a support of size $N$ \citep{Peyre}.
In this work, the regularized Wasserstein distance is calculated with a statistical procedure based on the Robbins-Monro algorithm for stochastic optimization. This way to calculate the Wasserstein distance is investigated in \citet{Stochastic_Genevay} and \citet{Stochastic_Bigot_Bercu}. The keystone of this approach is that the regularized Wasserstein problem can be written as the following stochastic optimization problem:
\begin{equation}\label{eq:expectation_wasserstein}
	W^{\varepsilon}(\alpha, \beta)=\max_{u \in \mathbb{R}^I} \mathbb{E}_{Y\sim\beta}[g_{\varepsilon, \alpha}(Y,u)],
\end{equation}
where $Y$ is a random variable with distribution $\beta$, and  $g_{\varepsilon, \alpha}(y, u)$ is easy to compute for any $y$ in $\beta$'s support and any $u \in \mathbb{R}^I$. The expression of $g_{\varepsilon, \alpha}$ can be found in equation \eqref{app-eq:in_the_expectation} of the Appendix. This formulation of the regularized Wasserstein distance as the maximum of an expectation allows the application of stochastic optimization methods. We stress that the large number of observations in cytometry data sets makes those stochastic approaches particularly relevant. For a more detailed presentation of stochastic algorithms for optimal transport, we refer the reader to Section \ref{app-sec:stocastic_regularized_wass} of the Appendix.

\subsection{Statistical model}

Let us consider $X_1^s,...,X_I^s$ the cytometry measurements from a first biological sample that is referred to as the source sample or source observations. The distribution of the source sample is modeled by a mixture of $K$ distributions: $\alpha = \sum_{k=1}^K \rho_k \alpha_k$,
where each term of the mixture corresponds to the cytometry measurements of one type of cell. For instance, $\alpha_k$ represents the underlying distribution behind the cytometry measurements of the CD4 effector T cells that are in the biological sample. The weights $\rho \in \Sigma_K$ are coefficients lying in the probability simplex $\Sigma_K = \{h \in \mathbb{R}_{+}^K ~:~ \sum_{k=1}^K h_k = 1\}$, and $\rho_k$ represents the proportion of one cell sub-type among all the cells found in the sample. 

We also consider a second set of cytometry measurements $X_1^t,...,X_J^t$ that is referred to as the target sample or target observations, and that corresponds to a second biological sample that may come from another patient or a cytometry analysis performed in a different laboratory. 
We assume that the underlying distribution of the the target observations is another mixture of $K$ distributions: $\beta = \sum_{k=1}^K \pi_k \beta_k$, where $\pi \in \Sigma_K$ represents the {\it} unknown class proportions and $\beta_k$ corresponds to the distribution of the $k^{\text{th}}$ cell population. Note that, in the framework of flow cytometry, we cannot make the assumption that $\alpha_k = \beta_k$, neither that $\rho_k = \pi_k$. Indeed, $\rho$ could differ from $\pi$ due to biological differences. For instance if the source sample comes from a healthy patient and the target sample comes from a sick patient, $\pi$ may feature significant differences from $\rho$. Moreover, technical variability of cytometry measurements could induce differences (e.g.\ location shift) between each component $\alpha_k$ and $\beta_k$ whereas these two distributions represent the same biological phenomenon.

These potential differences between the distributions of two flow cytometry measurements make the development of supervised methods to infer the unknown proportions $\pi$ in the target data from those of the source data a difficult task.
In this work, we rely on the geometric properties of the Wasserstein distance to handle the differences between samples that are due to technical reasons or inter-variability between healthy and sick patients.

\section{Class proportions estimation}\label{sec:method}

We now detail a supervised algorithm that estimates the class proportions $\pi$ in the target distribution. While state-of-the-art  automated methods in cytometry data analysis attempt to classify the observations $X_1^t,...,X_J^t$, we go straight to the estimation of the class proportions in the unsegmented target data set. To do so, we borrow ideas from the domain adaptation technique proposed in \citet{Target_shift} where it is proposed to re-weight the source observations by searching for weights minimizing the regularized Wasserstein distance between the re-weighted source measure and the target measure. Contrary to \citet{Target_shift}, we handle the regularized Wasserstein distance and the minimization problem by a stochastic approach. Indeed the large number of observations produced by flow cytometry makes stochastic techniques competitive to handle the high-dimensionality of such data sets.

\subsection{A new estimator of the class proportions}\label{subsec:def_estimator}

We now specify the definition of the estimator of the class proportions in the target data set. For the target sample $X_1^t,...,X_J^t$, the segmentation into various cell sub-types is not available, hence we define the empirical target measure as
$
\hat{\beta} = \frac{1}{J} \sum_{j=1}^{J} \delta_{X_j^t}
$.
From the source observations $X_1^s,...,X_I^s$, we define the empirical source measure as
$\hat{\alpha} = \frac{1}{I} \sum_{i=1}^I \delta_{X_i^s}$. Then, the knowledge of the gating of the source data allows to re-write the measure $\hat{\alpha}$ as a mixture of probability measures where each component corresponds to a known sub-population of cells in the data
\begin{equation}\label{Source_as_mixture}
	\hat{\alpha} = \sum_{k=1}^K \frac{n_k}{I} \left( \sum_{i: X_i^s \in C_k} \frac{1}{n_k} \delta_{X_i^s} \right) = \sum_{k=1}^K \frac{n_k}{I} \hat{\alpha}_k,
\end{equation}
where $n_k = \#C_k$ and $\hat{\alpha}_k = \sum_{i: X_i^s \in C_k} \frac{1}{n_k} \delta_{X_i^s}$. Namely, the component $\hat{\alpha}_k$ is the empirical measure of the observations that belong to the  sub-population $C_k$ (known class). Then, instead of only considering the true class proportions $(n_1/I,...,n_K/I)$ in the source data set, we can re-weight the clusters $C_k$ in the empirical distribution as desired. Indeed, for a probability vector $h = (h_1,...,h_K) \in \Sigma_K$ we can define the measure $\hat{\alpha(h)}$ that corresponds to the re-weighted measure $\hat{\alpha}$ such that for all $k \in \{1,...,K\}$ the component $\hat{\alpha}_k$ amounts for $h_k$ in the measure $\hat{\alpha}(h)$. In mathematical terms, the measure $\hat{\alpha}(h)$ is thus defined by:
\begin{equation}\label{eq:reweighted_source}
	\hat{\alpha}(h) = \sum_{k=1}^K h_k \hat{\alpha}_k.
\end{equation}
Then, to derive the class proportions in the target data, we minimize the regularized Wasserstein  distance \eqref{eq:regularized_problem} between the re-weighted source empirical distribution \eqref{eq:reweighted_source} and the target empirical distribution. 
The main idea is that the source distribution will get closer to the target distribution as the class proportions in its re-weighted version get closer to the class proportions of the target distribution. Thus, we propose to estimate the weights $\pi = (\pi_1,...,\pi_K) \in \Sigma_K $ of the underlying distribution $\beta$ behind the observations $X_1^t,...,X_J^t$ of the unlabelled target data set by
\begin{equation}\label{eq:estimator_proportion}
	\hat{\pi} \in \argmin_{h \in \Sigma_K} W^{\varepsilon}(\hat{\alpha}(h), \hat{\beta}).
\end{equation}
The combination of the estimator $\hat{\pi}$ and the algorithms described in Subsection \ref{subsec:minimization_strategies} to solve the associated minimization problem \eqref{eq:estimator_proportion} will be referred to as {\tt CytOpT}.

\subsection{A brief overview of the minimization procedure}\label{subsec:minimization_strategies}

The optimization problem \eqref{eq:estimator_proportion} leading to the estimator $\hat{\pi}$ of the class proportions does not have a solution in a closed form expression, and a numerical approximation is needed. We propose two methods to solve this optimization problem. In this section, we only give some insights of these two methods without focusing on the technical details. For a more thorough presentation, we refer the reader to the appendix. Our first strategy to tackle Problem \eqref{eq:estimator_proportion} consists in using a descent-ascent algorithm where the inner loop yields a stochastic approximate of $\nabla_h W^{\varepsilon}(\hat{\alpha}(h), \hat{\beta})$. For the second procedure, we borrowed ideas from \citet{ballu2020stochastic}. In this second strategy, we add a regularizing term to Problem \eqref{eq:estimator_proportion} that permits to swap the min and the max in order to rewrite this problem as a simple expectation maximization problem. Therefore, this regularized version of Problem \eqref{eq:estimator_proportion} could be solved with a straightforward stochastic gradient ascent. From our numerical experiments we have found that the second strategy is ten times faster than the descent-ascent procedure. As in practice both procedures seem to provide close estimates, all the results reported where produced with the second strategy, referred to as the "min-max swapping" strategy. 

\subsection{Some computational considerations}\label{subsec:computational_considerations}
The constant increase of the number of cellular markers used in cytometry analysis raises the question of the impact of such higher-dimensions, both for manual gating (dramatically increasing the resources necessary) and automated approaches (impacting the computational cost and efficiency). 
As mentioned earlier, Wasserstein metrics drag a computational burden.
But regarding \texttt{CytOpT}, the increase in the number of markers does not extend the computational cost by much. Indeed, the dimension only impacts the initial computation of the distance matrix between the observations.  Moreover, the random procedures allow a computational cost independent of the number of observations in the target data set. To be more specific, we study the evolution of the computational cost depending on the variables of interest that are; the number of markers measured $d$, the number of clusters $K$, and the number of observations $I$ in the source and in the target $J$.
For the descent ascent optimization strategy, which is a double loop algorithm, the computational cost is of order $O(n_{out}n_{in}IdK^2)$  where $n_{out}$ denotes the number of iterations of the outer loop and $n_{in}$ the number of iterations of the inner loop. In the case where the second strategy is preferred, a single loop algorithm can be applied at a computational cost of order $O(n_{iter}IdK)$. Here, $n_{iter}$ denotes the number of iterations of the stochastic gradient algorithm. Notice that in both cases, the random aspect of the optimization procedures provides a computational cost that is independent of the number of observations $J$ in the target data set.

\subsection{Extension of {\tt CytOpT} to a soft assignment method}\label{sec:label_propagation}

While our method aims at estimating the class proportions and not to classify the target observations, regularized optimal transport offers a natural soft assignment method which can be used to derive a soft classification of the target data set, as illustrated in Figure \ref{fig:transport_plan}. Assuming that we have access to the optimal transport plan $P_{\varepsilon}$ with respect to the regularized problem \eqref{eq:regularized_problem}, the coefficient $(P_{\varepsilon})_{i,j}/b_j$ can be interpreted as the probability that $X_i^s$ is assigned to $X_j^t$. Here, $b_j$ is the weight associated to the observation $X_j^t$. Thus, the probability $\gamma_j^{(k)}$ that $X_j^t$ belongs to the class $C_k$ is 
$\gamma_j^{(k)} = \frac{1}{b_j}\sum_{i=1}^I 1_{X_i^s \in C_k}(P_{\varepsilon})_{i,j}$
By choosing the class with highest probability, we can derive a classification for the observation $X_j^t$. To compute an approximate $\hat{P}$ of $P_{\varepsilon}$ one can plug $\widehat{U}$, the Robbins-Monro approximation of the optimal dual vector $u^*$, in formula \eqref{app-eq:dual_to_primal} of the supplementary. Thus, it is possible derive an automatic gating of the target data thanks to optimal transport. However, the transfer of the classification from the source data set toward the target data set requires to compute all the columns of the optimal transport plan. Therefore, obtaining a classification in the target data calls for additional calculations, while in clinical applications, the useful information is very often the relative proportions of the different cell sub-types. That is why this work focuses on the estimation of class proportions. 

\subsection{Measure of performance}\label{subsec:kullback_divergence}

To evaluate our approach on real flow cytometry data, the benchmark class proportions $\pi = (\pi_1,...,\pi_K) \in \Sigma_K $, will be defined thanks to the manual segmentation of the target observations. We recall that our algorithm does not make use of the segmentation of the target data set, and that it is only used to evaluate our method. In flow cytometry data analysis the $F$-measure is a popular tool to assess the performances of the clustering methods. As our method does not yield a clustering but an estimate $\hat{\pi}$ of the class proportions $\pi$, we cannot rely on this measure. Indeed, \CytOpT{} yields a probability vector $\hat{\pi} \in \Sigma_K$ that we wish to be the closest to the class proportions $\pi \in \Sigma_K$ of the target data set. Hence, a natural way to measure the discrepancy between $\hat{\pi}$ and $\pi$, is the Kullback-Leilbler divergence. Thus, with an estimation $\hat{\pi} = (\hat{\pi}_1,...,\hat{\pi}_K)$ and a benchmark $\pi=(\pi_1,...,\pi_K)$, to assess the quality of the estimator $\hat{\pi}$, we compute the Kullback-Leibler divergence $\KL$ defined as
$\KL(\hat{\pi}|\pi) = \sum_{k=1}^K \hat{\pi}_k\log\left(\frac{\hat{\pi}_k}{\pi_k}\right)$.\\

We also relied on graphical diagnoses from Bland-Altman plots (\cite{bland1986statistical}) to visually assess and compare the performance of \texttt{CytOpT} for class proportions estimation. Those diagnoses provide an overview of \texttt{CytOpT} behavior and performance on a collection of several cytometry data sets at once. For one target data set, a Bland-Altman plot compares the estimation $\hat{\pi} = (\hat{\pi}_1,...,\hat{\pi}_K)$ with the benchmark $\pi = (\pi_1,...,\pi_K)$ by plotting the difference  $\hat{\pi} - \pi$ against the mean $(\hat{\pi} + \pi)/2$. To simultaneously visualize the results from the analysis of two data sets,
one just adds on to the graph the points defined by $\hat{\pi}^{(2)} - \pi^{(2)}$ for the y-axis and $(\hat{\pi}^{(2)} + \pi^{(2)})/2$ for the x-axis (with $\hat{\pi}^{(2)}$ denoting the estimation on the second data set and $\pi^{(2)}$ the benchmark on that data set). This way, one can actually represent the results from the analysis of several data sets in a single plot. We also display on the Bland-Altman plots the mean of the difference with a solid horizontal line, and with  dashed horizontal lines, $\pm 1.96$ times the standard deviation of the difference. As we aim for $\hat{\pi}$ to be as close as possible to $\pi$, the closer to the $x$-axis the points are, the better the performance.

\section{Simulation Study}\label{sec:simulation_study}

\subsection{Comparison with classification methods when targeting the weights of a mixture}

\begin{figure}[!ht]
	\makebox[\textwidth][c]{\includegraphics[width=1.2\textwidth]{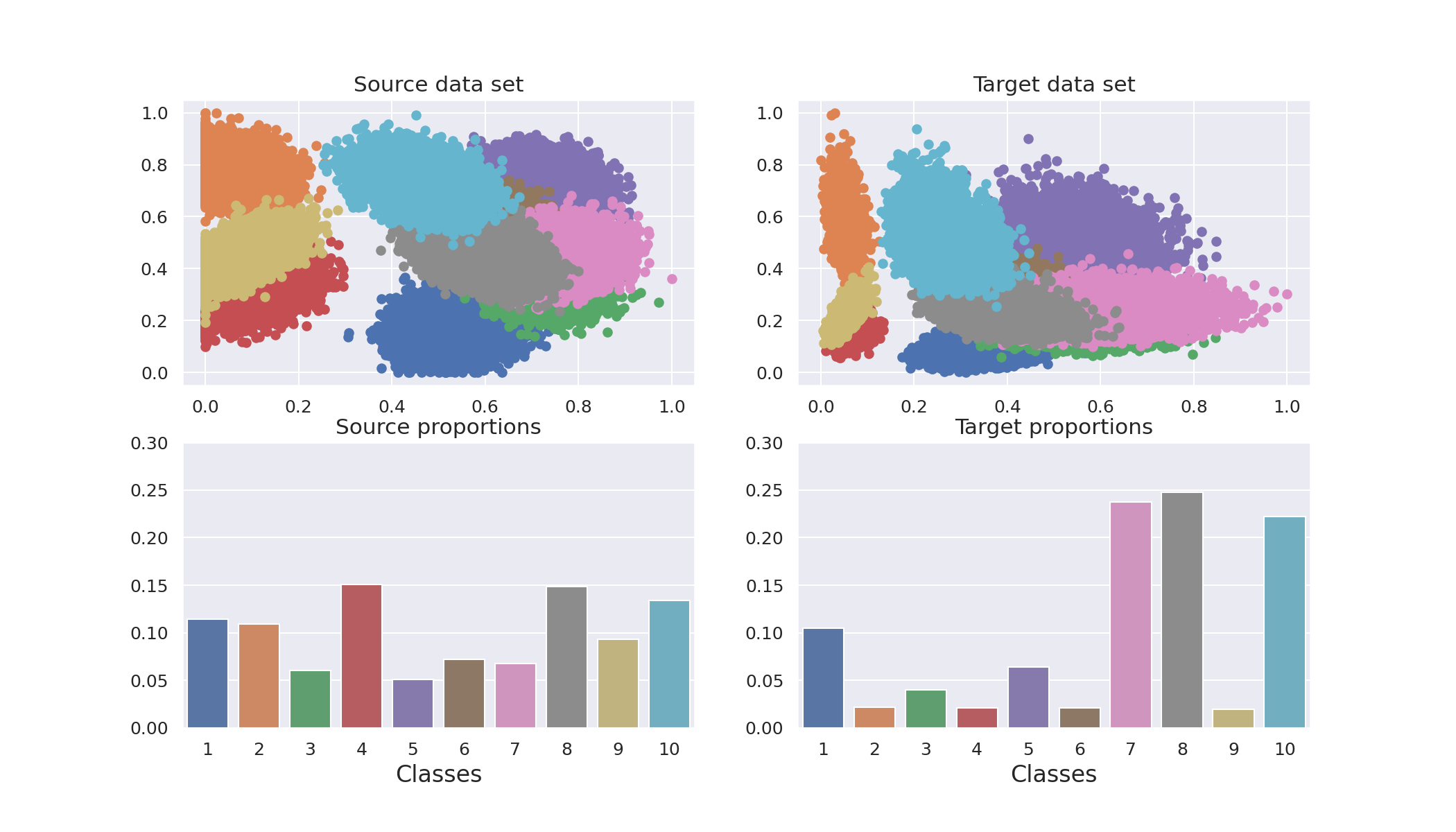}}    
	\caption{\footnotesize\textbf{Simulated source data set and Simulated target data set}. Using the segmentation of the source data set (Left) \texttt{CytOpt} estimates the class proportions in a target data set (Right) without making use of the classification of the target data set.}
	\label{fig:simulation_frame_work} 
\end{figure}

\begin{figure}[!ht]
	
	\makebox[\textwidth][c]{\includegraphics[width=1.2\textwidth]{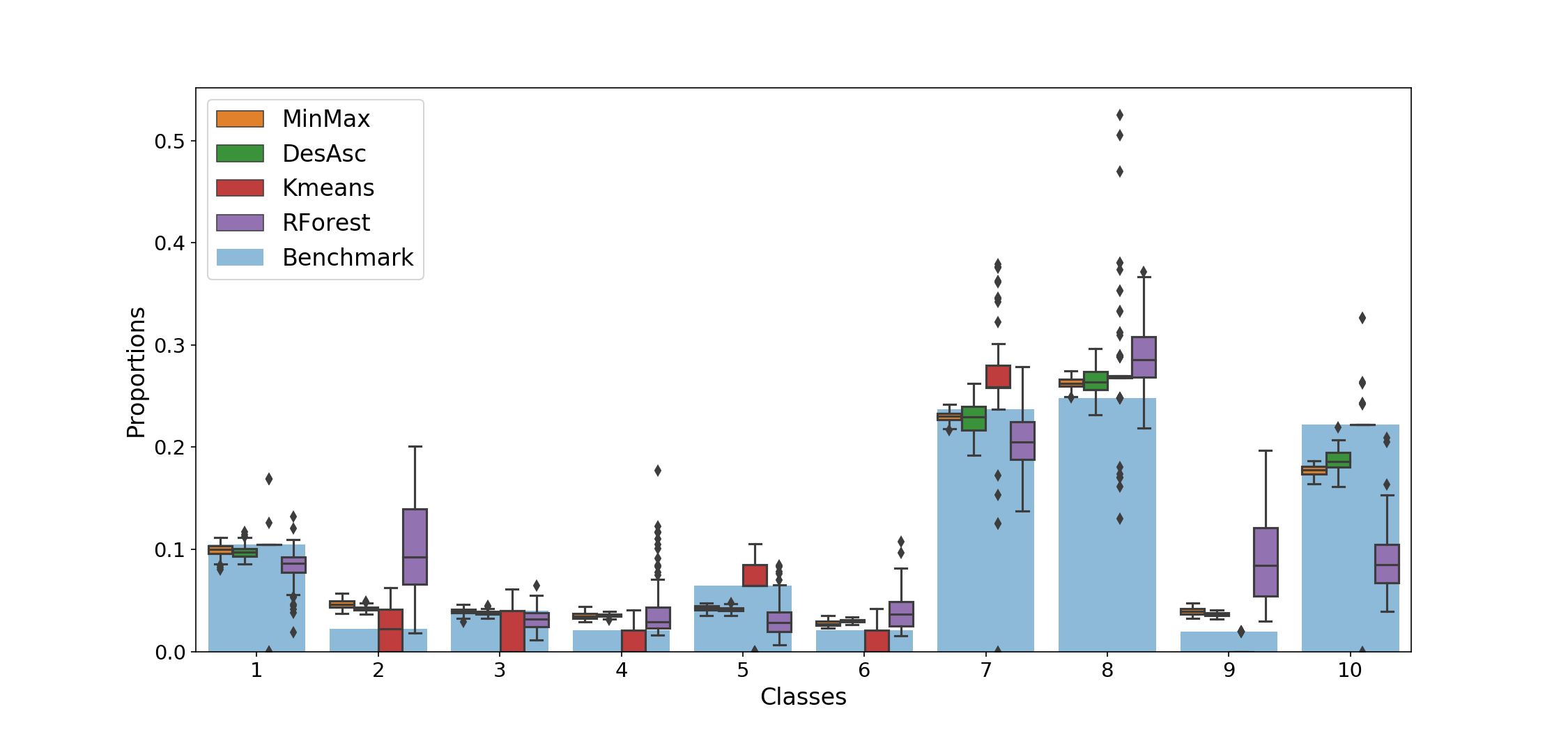}}	
	\caption{\footnotesize\textbf{Performances of the automated methods tested to estimate the class proportions}. For each method, one boxplot displays the median and the variation of the estimation of the weight of one component.}
	\label{fig:simulation_Boxplot} 
\end{figure}

To evaluate the performance of our method we generate two data sets from two different mixtures of K = 10 components each. The first data set is drawn from a Gaussian mixture distribution in dimension $d = 10$  with a vector of proportions $\rho \in \Sigma_K$. Thus, $\alpha = \sum_{k=1}^{10} \rho_k \alpha_k$ where the $\alpha_k$ are Gaussian distributions. The second data set is drawn from the same Gaussian components $\alpha_k$ but with different mixture proportions $\pi \neq \rho$. Then, we add a non-linear mapping $T : \mathbb{R}^d \to \mathbb{R}^d$ mimicking domain-shift. Thus, $\beta =  \sum_{k=1}^{10} \pi_k \beta_k$ where $\beta_k = T \# \alpha_k$ (that is the push-forward of the measure $\alpha_k$ by the mapping $T$). In this simulation study, we apply the same pre-processing of the data as the one applied to the real flow cytometry data. That is, we threshold the negative values at zero and re-scale the observations such that every observation $X$ belongs to $[0,1]^{10}$. See Figure \ref{fig:simulation_frame_work} for an illustration of a generated data-set. We compare our method with three state-of-the-art classification methods: one unsupervised method \--- the $k$-means ; and two supervised methods \--- quadratic discriminant analysis (QDA), and random forest (RF). After a training step on the source data set, each supervised method yields a classification of the target data set. Hence, after this classification step by either QDA or RF, we can derive an estimation of the class proportions in the target data set.  We indicate, that in this framework (namely with a single learning data set), the QDA algorithm corresponds to the \texttt{OptimalFlow} method \citep{Optimalflow}, that is a state-of-the-art method designed to cluster flow cytometry data.
As the $k$-means is an unsupervised method, we apply it straight to the target data set. Since the $k$-means does not yield a labelled classification, it is necessary to add an annotation step for each cluster $C_k$ returned by the $k$-means algorithm. To annotate a cluster $C_k$, we retrieve the majority population according to the true labels and then we label all the observations that belong to $C_k$ with this label. Therefore, we provide a significant upper hand to this method with this additional information. To study the performance and the stability of these various methods, we sampled 100 source data sets with respect to the source model (with a fixed number $I = 115,783$ of observations), and sampled 100 target data sets with respect to the target model  (with a fixed number $J = 68,981$ of observations). Then, for each pair of source and target data sets we apply the supervised methods, that is \texttt{CytOpT}, QDA and RF,  and on each target data set, we apply the $k$-means algorithm followed by the annotation step described above. Hence, for each method we obtain 100 estimates of the class proportions. 
Figure \ref{fig:simulation_frame_work} displays an example of a pair of source and target data sets with their corresponding class proportions.
Figure \ref{fig:simulation_Boxplot} displays the median class proportions retrieved with theses 100 estimates and the variability of these estimates. In this scenario, \texttt{CytOpT} is the method which offers the best trade-off between accuracy and stability. 

\subsection{Robustness evalauation}\label{subsec:robustness}

\begin{figure}[!ht]
	\makebox[\textwidth][c]{\includegraphics[width=1.2\textwidth]{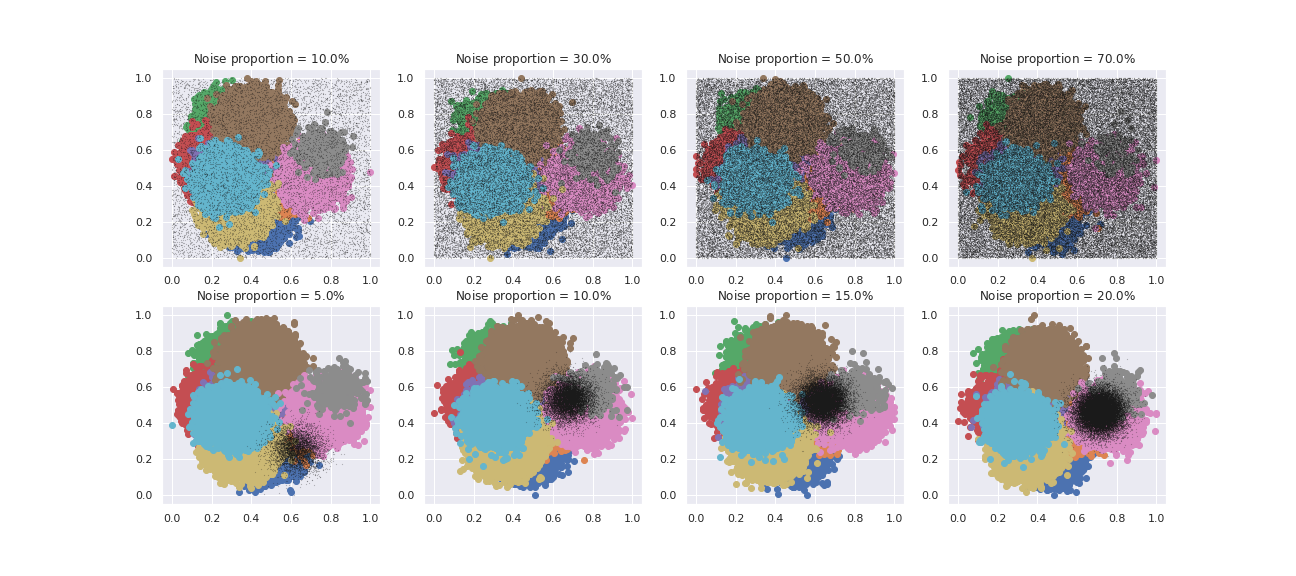}}	
	\caption{\footnotesize\textbf{Scenarios with additional observations in the target.} Top: the additional class (in black) is uniformly distributed in $[0,1]^d$. Bottom: the additional class (in black) is distributed with respect to a Normal distribution.}
	\label{fig:additionnal_data_target} 
\end{figure}

\begin{figure}[!ht]
	\makebox[\textwidth][c]{\includegraphics[width=1.2\textwidth]{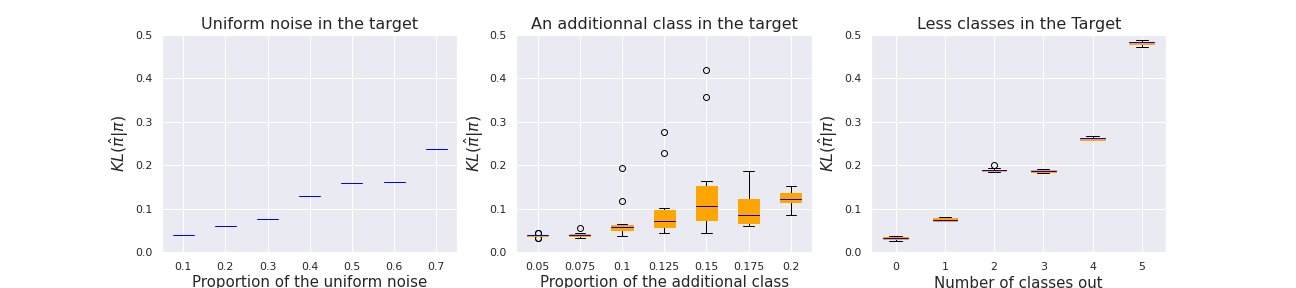}}	
	\caption{\footnotesize\textbf{Robustness assessment of CytOpt performance according to Kullback-Leibler divergence of the estimated proportion.}}
	\label{fig:results_robustness_assessment} 
\end{figure}

This section is devoted to the robustness analysis of \texttt{CytOpT}. To do so, we disrupted the ideal situation where the source mixture $\alpha = \sum_{k=1}^{10} \rho_k \mathcal{N}(\mu_k, \sigma I)$ and the target mixture $\beta = \sum_{k=1}^{10} \pi_k \mathcal{N}(\mu_k, \sigma I)$  share the same normal components but have different weights in three different ways. First, we added a uniform noise in $[0,1]^d$ to the target data set, to mimic outliers observations that can often be observed in practice. In our experiments, the noise amounted from $10 \%$ of the target data to $70\%$ of the target data. This first scenario is displayed in the upper panels of Figure \ref{fig:additionnal_data_target}. Second,  we added a $K+1^{th}$ class in the target data set that could not be found in the source data set. In this more involved situation, the additional class was also sampled from a Gaussian distribution with its center randomly chosen in $[0,1]^d$. This additional class represented up to $20\%$ of the target data set in some envisioned scenarios. This second scenario is displayed in the lower panels of Figure \ref{fig:additionnal_data_target}. Third, we remove from $1$ to $5$ components in the target mixture. Hence, in this last scenario more classes were presented in the source data set than in the target data set. 
For each of these three cases, we sampled $10$ couples (source data set, target data set) to get $10$ estimates of the class proportions for a given framework.
Figure \ref{fig:results_robustness_assessment} represent the performance of \texttt{CytOpt} in terms of Kullback-Leibler divergence in these three robustness trials. Unsurprisingly, the more we disrupt the initial setting, the less accurate the estimation. However, one can observe in Figure \ref{fig:results_robustness_assessment} that the increase of the Kullback-Leibler divergence is proportional to the number of classes removed or to the proportions of the noisy observations in the target. Those results are reassuring for practical applications as it shows that some reasonably mild disturbances in the target will not lead to meaningless results.

\section{Application to real flow cytometry data analysis}\label{sec:results}

\subsection{Illustration of the methodology with a two classes example}\label{subsec:2d_2classes}

In this section, we illustrate our proposed method in the setting where the cytometry data from HIPC described in Section \ref{subsec:data_set} are divided only among two broad classes: the CD4 cells and the CD8 cells. For ease of visualization, we use only two markers: the CD4 marker and the CD8 marker \--- that is $d=2$. This basic case, where two-dimensional data are divided into two classes, is a first illustration of our method in a favorable situation. We aim to demonstrate that our approach may reach a good approximation of the CD4 proportion and the CD8 proportion in a target cytometry data set with unknown gating.

\begin{figure}[!ht]
	
	\makebox[\textwidth][c]{\includegraphics[width=1\textwidth]{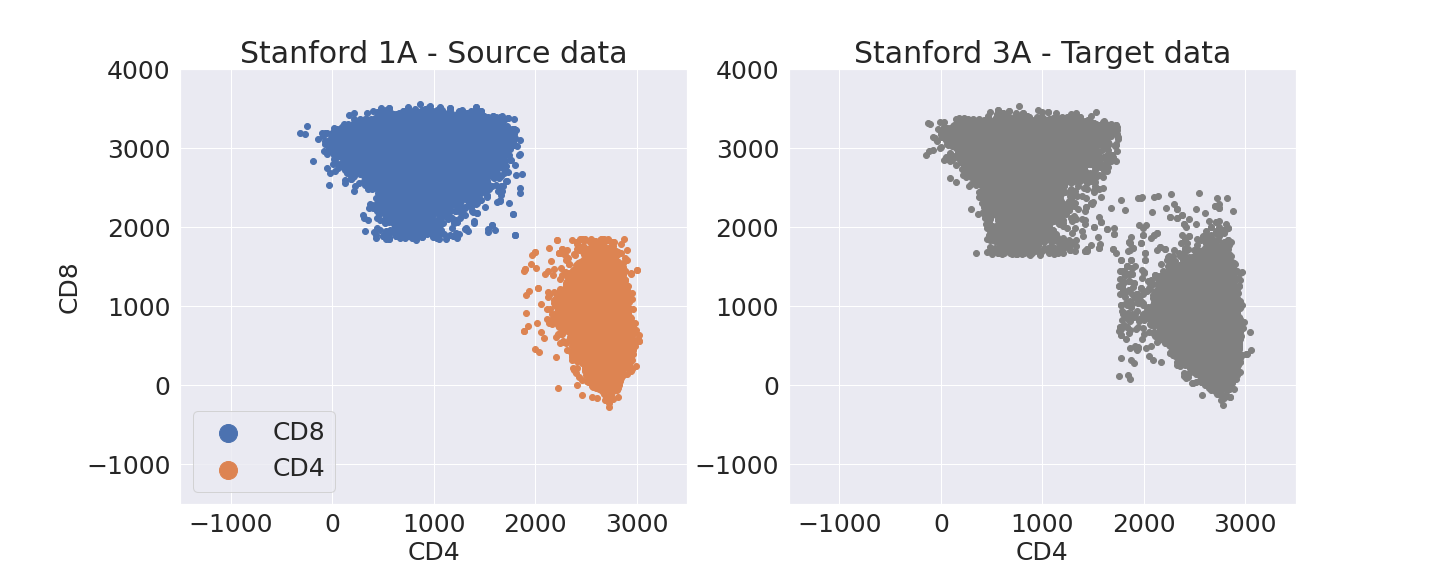}}	
	\caption{\footnotesize\textbf{Illustration of the \texttt{CytOpT} framework}. \texttt{CytOpt} estimates the class proportions in an unclassified data set (Stanford3A, right) from one classified data set (Stanford1A, left) without clustering the observations of Stanford3A.}
	\label{fig:frame_work_two_classes} 
\end{figure}

We consider the two data sets from the HIPC T-cell panel that are displayed in Figure \ref{fig:frame_work_two_classes}. The first data set is a series of cytometry measurements performed in a Stanford laboratory on a biological sample that comes from a patient identified as patient 1. This first data set is chosen as the segmented source data. The second data set is a series of cytometry measurements done in the same laboratory on a biological sample that comes from an other patient identified as patient 3. This second data set will be the target data set. The manual gating classification, and thus the class proportions are available for both samples. Nevertheless, we only use the classification available for the source (i.e. Stanford1A) data set in order to estimate the class proportions in the target (i.e. Stanford3A) data set. The manual gating classification for the target data set is only used as a benchmark to evaluate our method. 

First, to assess the relevance of our method, we present in Figure \ref{app-fig:wasserstein_evolution} the evolution of the regularized Wasserstein distance as a function of the weights associated to each class in the source distribution. To this end, we evaluate the function $F: h_1 \mapsto W^{\varepsilon}(\hat{\alpha}(h), \hat{\beta})$, where $h=(h_1, 1-h_1)$, on a finite grid $\mathcal{H} = \{ h^{(1)},...,h^{(m)} \}$. For $h_1 \in \mathcal{H}$ we approximate $W^{\varepsilon}(\hat{\alpha}(h), \hat{\beta})$ by the estimator $\widehat{W}_n$ defined in equation \eqref{app-eq:estimator_proportion}. It can be observed that the regularized Wasserstein distance decreases as the class proportions of the source data set get closer to the class proportions of the target data set. Using the segmentation of a cytometry data set, \texttt{CytOpT} adequately retrieves the true class proportions of an unlabelled cytometry data set. Even if the two classes situation is somewhat an easy scenario, one can observe the significant gap in the class proportions between the source data set and the target data set. In the Stanford1A data set the CD4 cells constitute 45.1 \% of the cells and the CD8 cells 54.9 \% of the cells, whereas in the Stanford3A data set the CD4 cells constitute 73.9 \% (estimated at 73.3 \% by \texttt{CytOpT}) of the cells and the CD8 cells 26.1 \% (estimated at 26.7 \% by \texttt{CytOpT}).

\begin{figure}[!ht]
	
	\makebox[\textwidth][c]{\includegraphics[width=1\textwidth]{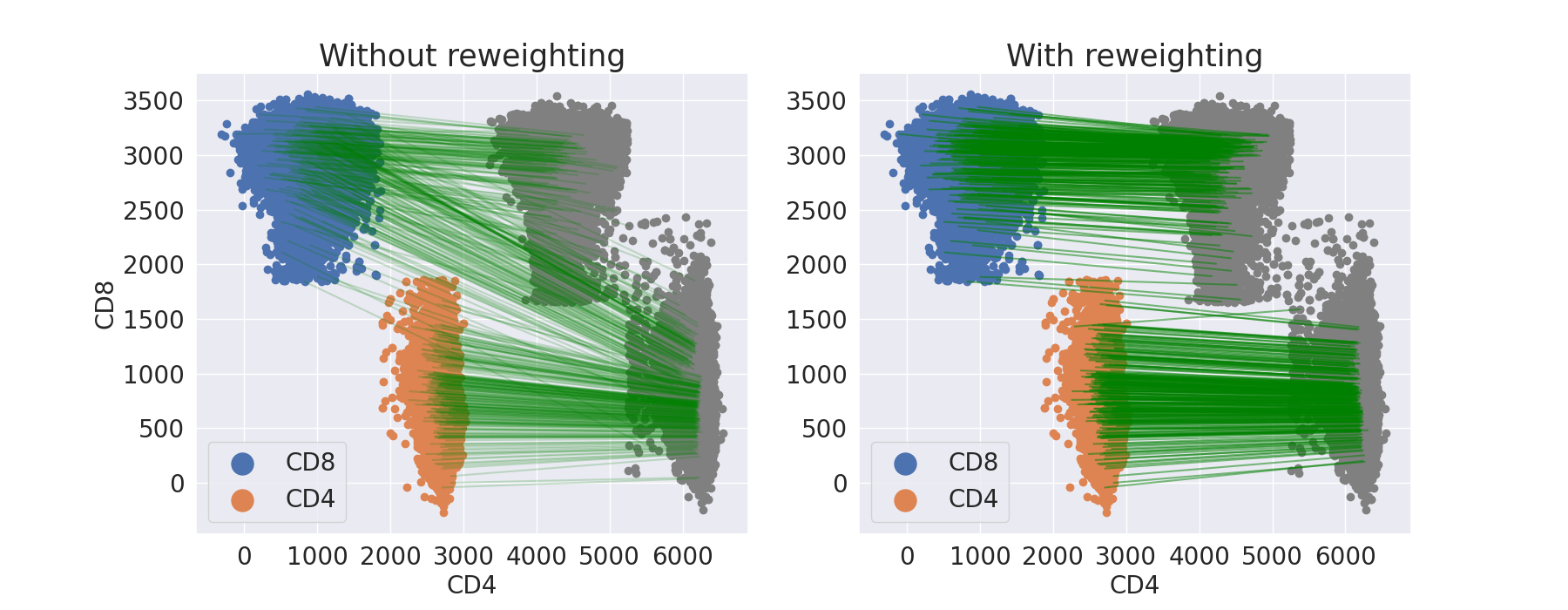}}    
	\caption{\footnotesize\textbf{Optimal transport plan $P_\varepsilon$ between the source and target distribution.} A green line between $X_i^s$ and $X_j^t$ indicates that the optimal transport plan $P_\varepsilon$ moves some mass from $X_i^s$ to $X_j^t$. To facilitate readability, the target data set has been shifted, and only 500 coefficients of $P_\varepsilon$ have been represented. Without re-weighting, mass from the CD8 class in the source data set is sent toward the CD4 class in the target data set. With re-weighting, the mass from one class in the source data set is sent to the corresponding class in the target data set.}
	\label{fig:transport_plan} 
\end{figure}

\begin{figure}[!ht]
	
	\makebox[\textwidth][c]{\includegraphics[width=1.2\textwidth]{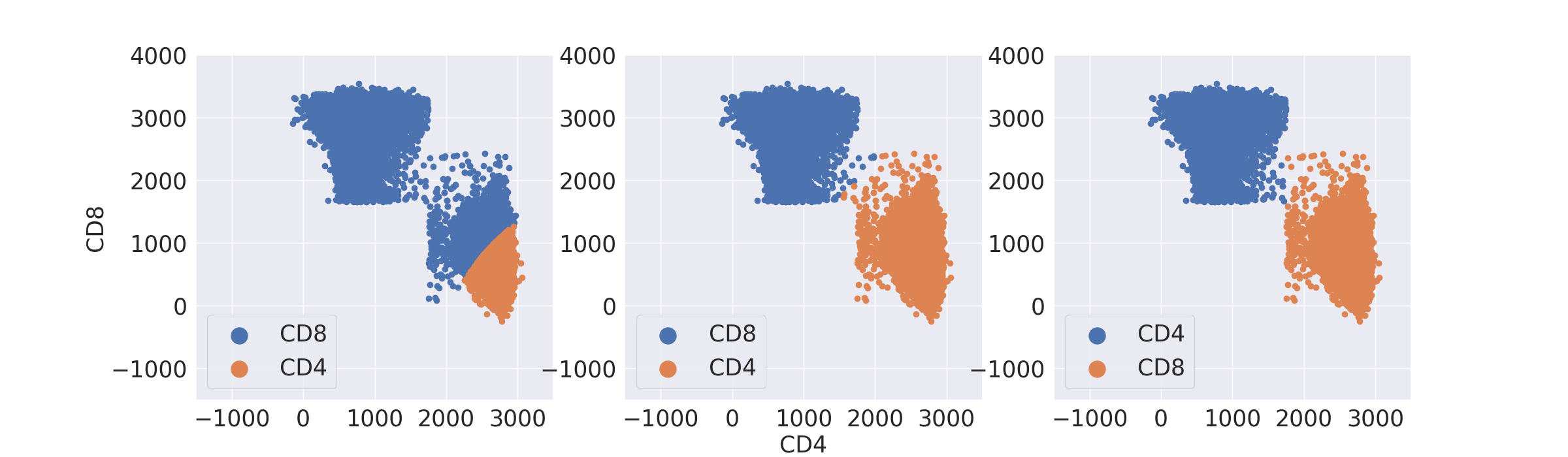}}	
	\caption{\footnotesize\textbf{Results of the soft assignment method and comparison with the manual clustering on the data set Stanford3A.} From left to right: Label transfer without reweighting - Label transfer with reweighting - Manual gating benchmark}
	\label{fig:res_label_propagation} 
\end{figure}

The results presented in Figure \ref{fig:transport_plan} and Figure \ref{fig:res_label_propagation} represent the use of the soft-clustering and classification methods described in Section \ref{sec:label_propagation}, and they outline the importance of re-weighting the source data. Without re-weighting, the classification obtained by optimal transport is really unlike the manual gating classification. However, one can re-weight the observations with the transformation $a(\pi) = \Gamma \pi $ in order to match the class proportions $\pi$ in the target data set. Here, $\Gamma$ is the linear operator defined in the equation \eqref{app-eq:class_to_observation} of the appendix. And once the source data are re-weighted with the estimated class proportions $\hat{\pi}$, the classification by regularized transport is very similar to the manual gating classification.

\subsection{Estimating 10 cell sub-type proportions from 7 cellular markers}

We apply our method to the full T-cell panel of the HIPC flow cytometry data described in Section \ref{subsec:data_set}. Contrary to the illustrative results presented in Section \ref{subsec:2d_2classes}, we now use all of the $d=7$ markers at hand to estimate the proportions of $K = 10$ cell populations. This represents a more involved classification task than the one considered in Section \ref{subsec:2d_2classes} with $K=2$. We arbitrarily chose the data set "Stanford1A" as the source measure to derive the class proportions in all the other data sets as targets.
A comprehensive evaluation of \texttt{CytOpT} performance is provided in Figure \ref{fig:bland_altman}, which features a Bland-Altman plot  \citep{bland1986statistical} displaying all the cell sub-types proportions estimations by \texttt{CytOpT} when targeting all available data sets across all the seven different centers available. In this example, we solely used the information from one reference classification from manual gating of one data set (namely Stanford1A) to estimate the class proportions in each of the 61 unsegmented data sets targeted.
In more than $90 \%$ of the cases the absolute difference between the estimated proportion and the manual gating gold-standard proportion is below $5 \%$. And in more than $99 \%$ of the cases the absolute error is no more than $10\%$.
Due to the stochastic nature of our algorithm, a new call to {\tt CytOpT} would lead to a slightly different estimate $\hat{\pi}$ compare to a former estimation. However, the results displayed in Figure \ref{fig:bland_altman} and Figure \ref{fig:barplot_estimation} are representative of the general quality of the estimation produced by {\tt CytOpT} in such settings.

Note that when applying our method to estimate class proportions on real flow cytometry data, a simple pre-processing is required. First, the signal processing of the cytometer can induce some contrived negative values of light intensity \citep{tung2007modern}. To undo this effect, we merely threshold those few negative values at zero. Second, to set the parameters of our algorithms, in particular $\varepsilon$, we need to bound the displacement cost. To do so, we scale the data such that: $\forall i \in \{1,...,I\},~ X_i^s \in [0,1]^d$ and $\forall j \in \{1,...,J\}, ~ X_j^t \in [0,1]^d$.

\subsubsection{Comparison with a straight application of one Manual
	Gating}\label{subsec:comparison_manual_gating}

\begin{figure}[!hp]
	\makebox[\textwidth][c]{\includegraphics[width=1.2\textwidth]{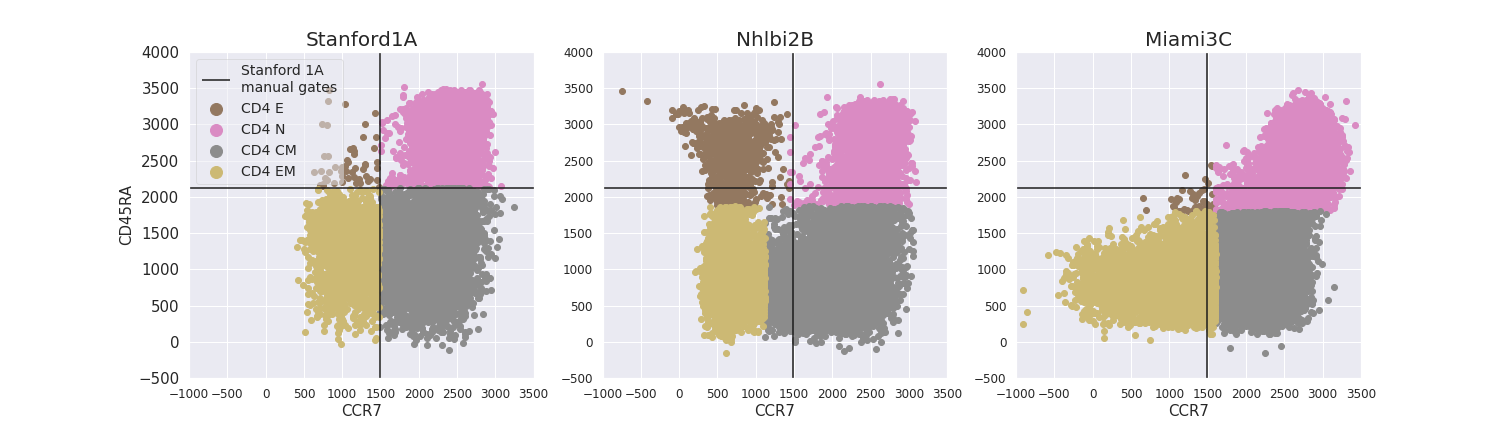}}    
	\caption{\footnotesize\textbf{2D projection of three cytometry data sets.} Left: CD4 cell measurements performed in Stanford. Middle: CD4 cell measurements performed in Nhlbi. Right: CD4 cell measurements performed in Miami.}
	\label{fig:Spatial_Shift_CD4} 
\end{figure}

\begin{figure}[!hp]
	\makebox[\textwidth][c]{\includegraphics[width=0.8\textwidth]{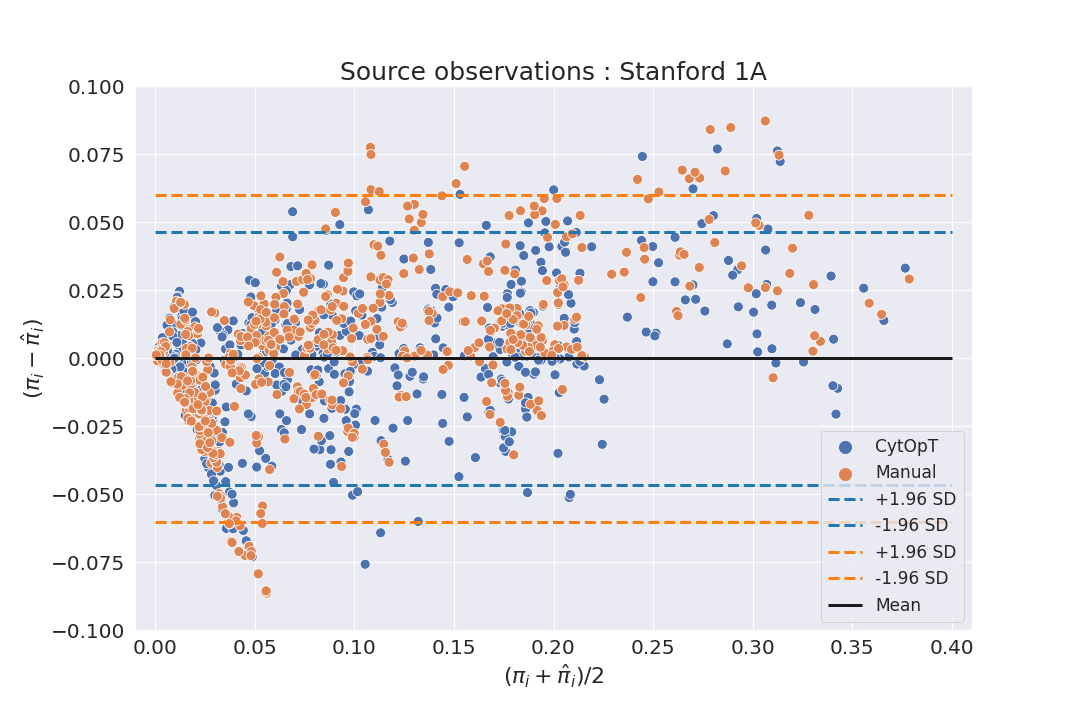}}    
	\caption{\footnotesize\textbf{Bland-Altman of \texttt{CytOpT} and Stanford1A's gating compared to specific manual gating benchmark.}}
	\label{fig:Bland_Altman_Manual_Gating} 
\end{figure}

As previously stated, technical variability across samples can lead to misalignment of flow cytometry data. For instance, spatial shift is displayed on Figure \ref{fig:Spatial_Shift_CD4} with only CD4 cells and the corresponding manual gating established on the basis of the Stanford1A measurements alone. While the four classes displayed on Figure \ref{fig:Spatial_Shift_CD4} represent the same biological phenomenon, it appears that the boundaries differ among the different centers. To illustrate the interest of \texttt{CytOpT} when technical variability induces a shift between the source and the target data set we compare \texttt{CytOpT} with the mere application of the Stanford 1A manual gates to the ungated data set. This comparison is performed on the 61 remaining ungated data sets. The bland-Alman displayed on Figure \ref{fig:Bland_Altman_Manual_Gating} shows that \texttt{CytOpT} provides a significant improvement over the direct application of Stanford1A's gating. We will now compare \texttt{CytOpT} with state-of-the-art automated approaches designed to analyze flow cytometry data.

\subsubsection{Comparison with other methods}

On the HIPC data, we compare \texttt{CytOpT} with 3 state-of-the-art automated-gating approaches specifically designed to analyze flow cytometry data. First, \texttt{flowMeans} \citep{aghaeepour2011rapid} which is a variation of the $k$-Means algorithm designed to cluster flow cytometry data. As \texttt{flowMeans} is an unsupervised method, we will label every observation of a cluster retrieved by \texttt{flowMeans} with the major population according to the reference manual clustering (Note that this in theory should provide a substantial advantage over the other methods). We also apply \texttt{Cytometree} to cluster the HIPC data. This method offers an automated annotation of its estimated clustering. Therefore, with \texttt{Cytometree}, we can directly retrieve an estimation of the class proportions. Finally, we compare the performances of our method with \texttt{OptimalFlow}, recently developed by \citet{Optimalflow} and which performs supervised automatic gating. We mention that it is the first method to make use of the Wasserstein distance in the context of automatic-gating of flow cytometry data analysis. We stress that, even though both our method and \texttt{OptimalFlow} are supervised approaches relying on the Wasserstein distance, they greatly differ. First, they have different objectives: \texttt{OptimalFlow} aims at classifying the cells, while \texttt{CytOpT} directly aims at estimating the cellular sub-types proportions. Second, \texttt{OptimalFlow} uses the Wasserstein distance to cluster a compendium of pre-gated cytometry data sets and uses the notion of Wasserstein barycenter to produce a prototype data set for each cluster. Then, using the Wasserstein distance, \texttt{OptimalFlow} browses among the prototypes to extract the most relevant prototype to classify a test data set. But ultimately, the classification is performed with tools such as \texttt{tclust} \citep{dost2010tclust}, or Quadratic Discriminant Analysis \citep{hastie2009elements} that do not belong to the field of optimal transport. On the contrary, \texttt{CytOpT} only requires one pre-gated data set and the estimator of the class proportions is fully based on regularized optimal transport.
In this comparison, \OptimalFlow{} had a learning database composed of the nine Stanford cytometry data sets, and it was set to produce one prototype from the database (Note that \OptimalFlow{} thus leverages the information from multiple source data sets, not just one). For the rest of the method, we applied \OptimalFlow{} with the default settings proposed in the vignette\footnote{ at \url{https://github.com/HristoInouzhe/optimalFlow}}. 

\begin{figure}[!hp]
	\makebox[\textwidth][c]{\includegraphics[width=1\textwidth]{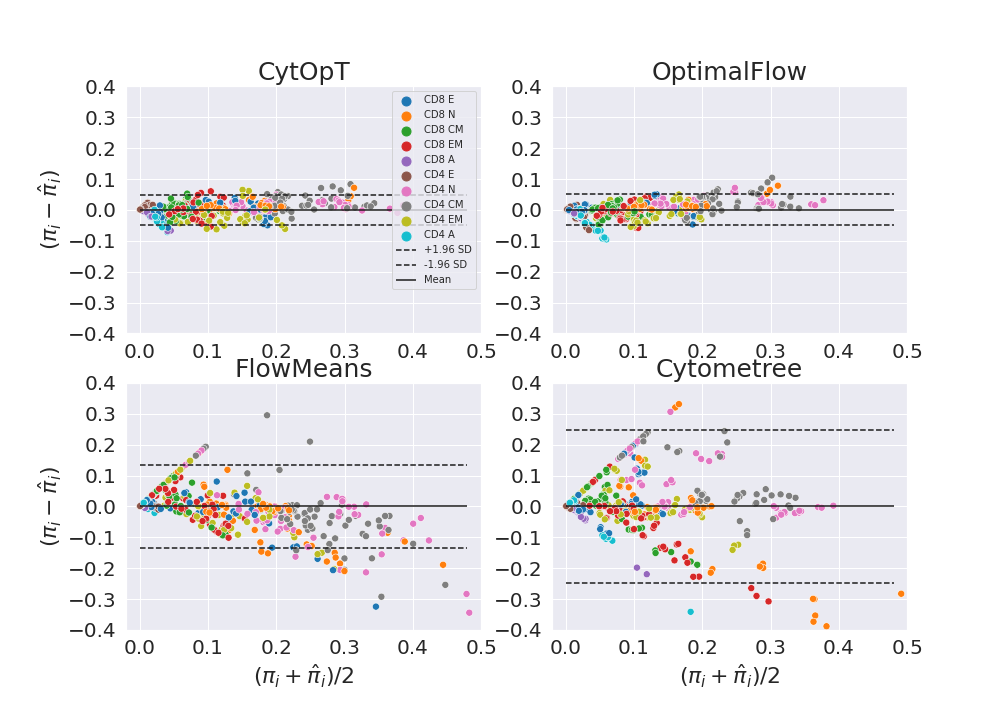}}
	\caption{\footnotesize\textbf{Comparison of the proportions $\hat{\pi}$ estimated by four automated methods and the manual benchmark $\pi$ on the HIPC database.} By plotting the difference against the mean of the results of two different methods, the Bland-Altman plot allows to assess the agreement between an automated method and the manual benchmark.}
	\label{fig:bland_altman} 
\end{figure}

Figure \ref{fig:bland_altman} displays a Bland-Altman plot for each of the compared method. \texttt{flowMeans} and \texttt{Cytometree} clearly stand-out with less accurate class proportions estimation that cannot compete with the good performance of \texttt{CytOpT} and \texttt{OptimalFlow}. Although visually \texttt{CytOpT} and \texttt{OptimalFlow} seem to yield close results, one can notice that \texttt{OptimalFlow} estimations are systematically biased as it underestimates the rarest cell populations proportions while overestimating the most represented cell populations frequencies.

\begin{figure}[!ht]
	\makebox[\textwidth][c]{\includegraphics[width=1\textwidth]{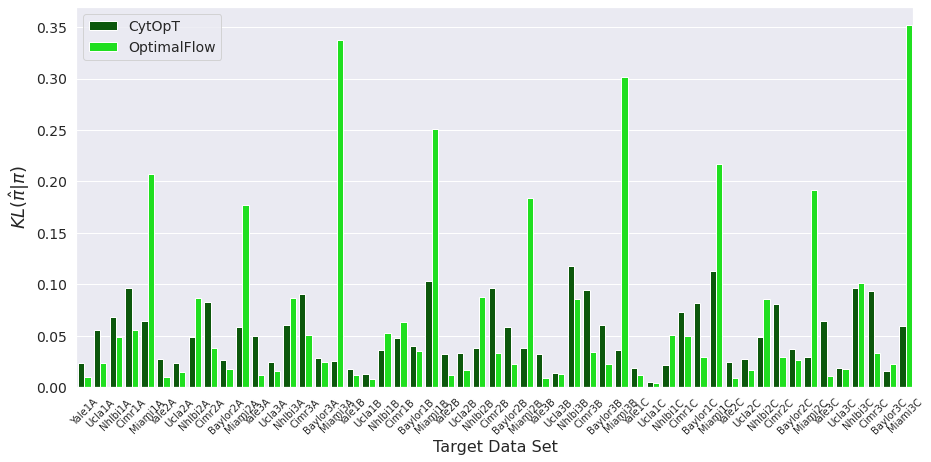}}
	\caption{\footnotesize \textbf{Comparison between our algorithm \texttt{CytOpT} and one of the state of the art automated method for cytometry data analysis: \texttt{OptimalFlow}.} The comparison was performed on the data sets of the HIPC panel.}
	\label{fig:comparison_methods} 
\end{figure}

\begin{figure}[!ht]
	\makebox[\textwidth][c]{\includegraphics[width=1\textwidth]{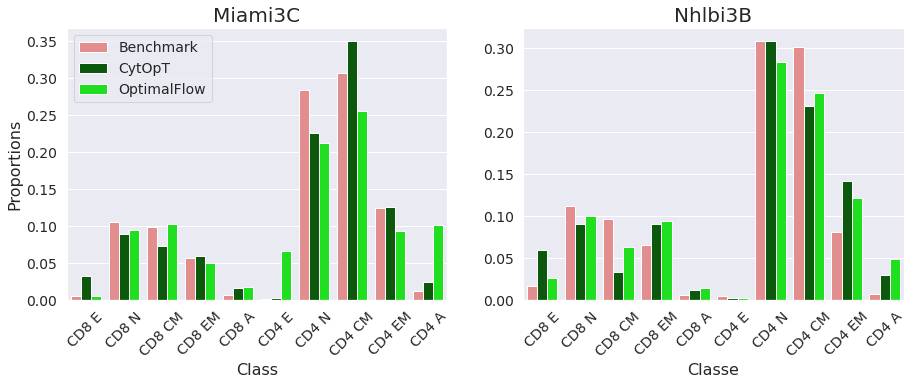}} 
	\caption{\footnotesize\textbf{Comparison of the estimated proportions $\hat{\pi}$ by \texttt{CytOpT} and \texttt{OptimalFlow} with the manual gating benchmark $\pi$.} Left: The target data set is Miami3C. Right: The target data set is Nhlbi3B. In both cases, the data set Stanford1A has been used as a source data set for \texttt{CytOpT}.}
	\label{fig:barplot_estimation}  
\end{figure}

To get a more quantitative criterion to compare \texttt{CytOpT} and \texttt{Optimalflow}, we used the Kullback-Leibler divergence to assess the discrepancy between the gold-standard class proportions derived manually and the estimations of the two methods. Figure \ref{fig:comparison_methods} shows that on almost every data set the quality of estimation is rather close. However when the target data set comes from Miami, the Kullback-Leibler divergence between the benchmark and \texttt{CytOpT} estimations is significantly lower than the Kullback-Leibler divergence between the benchmark and \texttt{OptimalFlow} estimations. This means that in theses cases \texttt{CytOpT} performs significantly better than \texttt{OptimalFlow}. We account for theses results by the spatial shift between the source and the target. To end this comparison, we present in Figure \ref{fig:barplot_estimation} a comparison of the class proportions estimation between \texttt{CytOpT} and \texttt{OptimalFlow} in two cases. The first case is the worst estimation for \texttt{Optimalflow}, namely, when the target data set is Miami3C. In this case, \texttt{OptimalFlow} overestimates very much the proportion of CD4 effector and the proportion of CD4 activated. The second case corresponds to the worst estimation for \texttt{CytOpT}, namely when the target data set is Nhlbi3B. If \texttt{OptimalFlow} performs slightly better in this case, \texttt{CytOpT} still offers an acceptable and competitive estimation.

\subsection{Application to the OptimalFlow data}

\begin{figure}[!hp]
	\makebox[\textwidth][c]{\includegraphics[width=\textwidth]{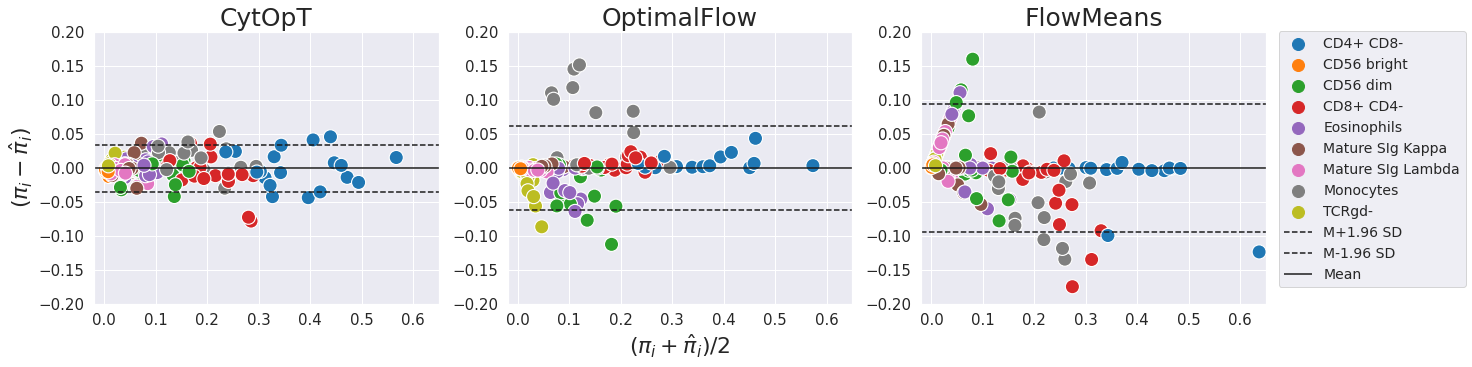}} 	
	\caption{\footnotesize\textbf{Comparison of the proportions $\hat{\pi}$ estimated with \texttt{CytOpT} and the manual benchmark $\pi$ on the OptimalFlow database.}}
	\label{fig:bland_altman_OFdata} 
\end{figure}

A second evaluation of our method was performed on the cytometry data used in \citet{Optimalflow}. This second cytometry panel is available on GitHub \footnote{ at \url{https://github.com/HristoInouzhe/optimalFlowData}}, and we refer to this panel as the \texttt{OptimalFlow} data sets. First, in order to run a sensible comparison between \texttt{CytOpT}, \texttt{OptimalFlow} and \texttt{FlowMeans} we selected nine cell populations that were present in the first 21 data sets. As \texttt{FlowMeans} is an unsupervised method, it was directly applied to the 21 data sets. For \texttt{OptimalFlow}, it was set to use the 3 first data sets of the database, to build one template, and then to apply the classification step. Hence \texttt{OptimalFlow} yielded a class proportions estimation for the 18 remaining data sets. For \texttt{CytOpT}, we used the first data set as a source data set and we estimated the class proportions in the 20 remaining data sets. Figure \ref{fig:bland_altman_OFdata} shows the Bland-Altman plots of the results for those three methods. One can notice that \texttt{CytOpT} is the method where the estimation of the class proportions $\hat{\pi}_k$ is overall the closest to the manual gold-standard $\pi_k$. \\

\section{Discussion}

We have introduced \texttt{CytOpT}, a supervised method that differs from existing method as it directly estimates the cell population proportions, without a preliminary clustering or classification step. While obtaining a classification of the observation is often a mathematical tool for achieving automatic gating, we emphasize that from a clinical perspective, the proportions of the different cell populations is the quantity of interest \citep{henel2007basic}. Thus, we have proposed efficient numerical schemes that rely on stochastic algorithms to address the delicate issue of computing and minimizing the regularized Wasserstein distance. We have demonstrated empirically, both on simulated data and on real flow cytometry data, the outstanding performance of \texttt{CytOpT} to tackle this question of cell population proportion estimation, especially in difficult situations where the technical variability of flow cytometry induces spatial shifts between samples. We also think that with higher-dimensional data, our method could perform even better as the Wasserstein distance would take advantage of all the dimensions available at once.

We present a few perspectives to further improve the performance of \texttt{CytOpT}. First, an additional pre-processing step to handle outliers could make \texttt{CytOpT} more robust to extreme cellular observations that have little biological meaning. Indeed, due to the needed scaling standardization of the data, one outlier observation can change the layout of the source data in comparison with the layout of the target data. A data driven strategy to choose the regularization parameter $\varepsilon$ could also help to tackle this outlier issue. Following the supervised approach of \citet{Optimalflow}, one could also propose an extension of \texttt{CytOpT} that would leverage the information from several labeled source data sets. A potential approach would be to select the most relevant labeled data set to estimate the class proportions in an unclassified target data set. To be more specific, let us assume that we have $M$ classified data sets $X^1,...,X^M$. For each data set $X^m$, we can define the re-weighted empirical measure $\alpha^m(h)$  such that the weight of the $k^{th}$ class equals $h_k$ for all $k \in \{1,...,K\}$, which would yield an optimal re-weighting $\hat{\pi}^m$ that corresponds to an estimation of the class proportions $\pi$ in the target data set. Finally, to select the most relevant estimator $\hat{\pi}^m$, we choose the one leading to the smallest Wasserstein cost between the re-weighted empirical measure $\hat{\alpha}^m(\hat{\pi}^m)$ and the empirical target distribution $\hat{\beta}$ that is :

\begin{equation}\label{eq:multi_source_estimate}
	\hat{\pi} = \argmin_{\hat{\pi}^1,...,\hat{\pi}^M}\{W^\varepsilon(\hat{\alpha}^1(\hat{\pi}^1), \hat{\beta}),..., W^\varepsilon(\hat{\alpha}^M(\hat{\pi}^M), \hat{\beta}) \}
\end{equation}
Another extension of this work could also consider the Sinkhorn divergence $S^{\varepsilon}(\alpha, \beta) =$ $W^{\varepsilon}(\alpha, \beta) - \frac{1}{2}W^{\varepsilon}(\alpha, \alpha) - \frac{1}{2} W^{\varepsilon}(\beta, \beta)$, instead of $W^{\varepsilon}(\alpha, \beta)$. Indeed, the authors of \citet{MMD_Sinkhorn} showed that $S^{\varepsilon}$ has theoretical properties that make it a suitable loss for applications in machine learning.

\section*{Codes and data availability}

The illustrative HIPC data sets described in Subsection \ref{subsec:data_set}, as well as Python Notebooks to reproduce the Figures presented in this paper are available at \url{https://github.com/Paul-Freulon/CytOpt}. A user-friendly Python package is available from pypi with source code at \url{https://pypi.org/project/CytOpT/}. A user-friendly R package, relying on the reticulate framework \citep{reticulate2022} to encapsulate the above Python code, will be available from CRAN with source code at \url{https://github.com/sistm/CytOpt-R}.

\section*{Acknowledgments}

The authors would like to thank the associate editor and the three referees for their suggestions and constructive comments
which helped to improve the paper substantially. The author would also like to thank Kalidou Ba for developing the \texttt{CytOpT} packages.
Experiments presented in this paper were carried out using the PlaFRIM experimental testbed, supported by Inria, CNRS (LABRI and IMB), Université de Bordeaux, Bordeaux INP and Conseil Régional d’Aquitaine.
J\'er\'emie Bigot is a member of Institut Universitaire de France (IUF), and this work has been carried out with financial support from the IUF.

\clearpage

\appendix

\renewcommand\thefigure{\thesection.\arabic{figure}}
\setcounter{figure}{0}
\renewcommand\theequation{\thesection.\arabic{equation}}
\setcounter{equation}{0}    

\section{Stochastic algorithm for the regularized Wasserstein distance}
\label{app-sec:stocastic_regularized_wass}

In this work, the regularized Wasserstein distance is calculated with a statistical procedure based on the Robbins-Monro algorithm for stochastic optimization. This way to calculate the Wasserstein distance is investigated in \cite{Stochastic_Genevay} and \cite{Stochastic_Bigot_Bercu}. The keystone of this approach is that the regularized Wasserstein problem can be written as the following stochastic optimization problem:
\begin{equation}\label{app-eq:expectation_wasserstein}
	W^{\varepsilon}(\alpha, \beta)=\max_{u \in \mathbb{R}^I} \mathbb{E}[g_{\varepsilon, \alpha}(Y,u)],
\end{equation}
where $Y$ is a random variable with distribution $\beta$, and  $g_{\varepsilon, \alpha}$ is defined as:
\begin{equation}
	g_{\varepsilon, \alpha}(y,u) = \sum_{i=1}^I u_i a_i + u_{c,\varepsilon}(y) - \varepsilon, \label{eq:geps}
\end{equation}
where $u_{c,\varepsilon}(y_j) = \varepsilon \left(\log(b_j) - \log \left(\sum_{i=1}^{I} \exp \left(\frac{u_i - c(x_i, y_j)}{\varepsilon} \right)\right) \right).$
We point out that the solution $P_{\varepsilon}$ of the regularized primal problem \eqref{eq:regularized_problem} is recovered from any $u^{*} \in \mathbb{R}^I$ solution of the semi-dual problem \eqref{eq:expectation_wasserstein} as:
\begin{equation}\label{app-eq:dual_to_primal}
	(P_{\varepsilon})_{i,j} = \exp\left( \frac{u_i^* + u_{c, \varepsilon}^*(y_j) - c(x_i, y_j)}{\varepsilon} \right).
\end{equation}
Formulation \eqref{app-eq:expectation_wasserstein} and the fact that for all $y \in \mathbb{R}^d$, the function $g(y,.)$ is concave lead us to estimate the vector $u^{*}$ by the Robbins-Monro algorithm \cite{Robbins_Monro} given, for all $n \geq 0$, by:
\begin{equation}\label{eq:robbins_monro}
	\widehat{U}_{n+1} = \widehat{U}_{n} + \gamma_{n+1}\nabla_{u}g_{\varepsilon}(Y_{n+1}, \widehat{U}_n),
\end{equation}
where the initial value $\hat{U}_{0}$ is a random vector which can be arbitrarily chosen, $Y_1,...,Y_{n+1}$ is an i.i.d. sequence of random variables sampled from the distribution $\beta$, and $(\gamma_n)_{n \geq 0}$ is a positive sequence of real numbers decreasing toward zero satisfying 
\begin{equation}\label{app-eq:step_size_condition}
	\sum_{n=1}^{\infty} \gamma_n = +\infty \qquad \text{and} \qquad \sum_{n=1}^{\infty} \gamma_n^2 < +\infty.
\end{equation}
It follows from \cite{Lightspeed} that the maximizer $u^{*}$ of \eqref{app-eq:expectation_wasserstein} is unique up to a scalar translation of the form $u^{*} - t1_{I}$ for any $t \in \mathbb{R}$. Throughout this paper, we shall denote by $u^{*}$ the maximizer of \eqref{app-eq:expectation_wasserstein} satisfying $\langle u^{*}, 1_I \rangle = 0 $ which means that $u^{*}$ belongs to $\langle 1_J \rangle^{\perp}$ where $\langle 1_J \rangle$ is the one-dimensional subspace of $\mathbb{R}^I$ spanned by $1_J$. As shown in \cite{Stochastic_Bigot_Bercu}, to obtain a consistent estimator of $u^*$, one must slightly modify the Robbins-Monro algorithm by requiring that $\widehat{U}_0$ belongs to $\langle 1_I \rangle^{\perp}$. To approximate the regularized Wasserstein distance $W^{\varepsilon}(\alpha, \beta)$ \cite{Stochastic_Bigot_Bercu} proposed the recursive estimator:
\begin{equation}\label{app-eq:estimate_wasserstein}
	\widehat{W}_n = \frac{1}{n} \sum_{k=1}^n g_{\varepsilon}(Y_k, \widehat{U}_{k-1}).
\end{equation}
Throughout this work, we will use this estimator.

\section{Two different numerical optimization schemes}\label{app-seq:solving_in_detail}

In this supplementary to the article, we present the technical details of the optimization strategies to compute the estimator of the class proportions which is given by:
\begin{equation}\label{app-eq:estimator_proportion}
	\hat{\pi} \in \argmin_{h \in \Sigma_K} W^{\varepsilon}(\hat{\alpha}(h), \hat{\beta}).
\end{equation}

In this optimization problem \eqref{app-eq:estimator_proportion} that yields an estimator $\hat{\pi}$ of the class proportions, the weights of the source distribution are modified, but the support is fixed. Thus, the parameter $h$ only impacts the reweighted measure $\hat{\alpha}(h)$ through its weights $a(h) = (a_1(h),...,a_I(h))$. We therefore introduce the linear operator $\Gamma \in \mathbb{R}^{I \times K}$ that maps the class proportions $h$ to the weight vector $a(h)$ with the constraint that among a given class $C_k$ every observation have the same weight $\frac{h_k}{n_k}$, where $n_k = \#C_k$.  This linear operator $\Gamma$ is defined by:
\begin{equation}\label{app-eq:class_to_observation}
	\forall (i,k) \in \{1,...,I\} \times \{1,...,K\}, ~ \Gamma_{i,k} =  \left\{
	\begin{array}{ll}
		\frac{1}{n_k} & \mbox{if } X_i^s \in C_k, \\
		0 & \mbox{otherwise.}
	\end{array}
	\right.
\end{equation}

Thanks to this transformation, the vector $a(h) = \Gamma h $ is such that the class proportions in the discrete measure $\sum_{i=1}^I a_i(h) \delta_{X_i^s}$ equals the vector $h$. 

\subsection{Descent-Ascent procedure}\label{app-subsec:first_method}

\paragraph{Formulation of the Problem}
To present the ideas of our first way to numerically solve $\eqref{eq:estimator_proportion}$, we re-write this optimization problem as:
\begin{equation}\label{app-eq:min_max_expectation}
	\min_{h \in \Sigma_K} W^{\varepsilon}(\hat{\alpha}(h), \hat{\beta}) = \min_{h \in \Sigma_K} \max_{u \in \mathbb{R}^I} \mathbb{E}[g_{\varepsilon, \alpha}(Y,u,h)],
\end{equation}
where $Y$ is a random variable with distribution $\hat{\beta}$, and for $y_j \in \mathbb{R}^d$ an observation of the random variable $Y$,
\begin{equation}\label{app-eq:in_the_expectation}
	g_{\varepsilon, \alpha}(y_j,u,h) = \sum_{i=1}^{I} u_i a_i(h) + \varepsilon \left(\log(b_j) - \log \left(\sum_{i=1}^{I} \exp \left(\frac{u_i - c(x_i, y_j)}{\varepsilon} \right)\right) \right) - \varepsilon.
\end{equation}

We mention that contrary to expression \ref{eq:geps}, the function $g_{\varepsilon, \alpha}$ that appears in expression \ref{app-eq:in_the_expectation} only depends of $\alpha$ through its support $X^s=(X_1^s,...,X_I^s)$. For ease of notation we stick to the notation $g_{\varepsilon, \alpha}.$.\\

Due to the min-max formulation of Problem \eqref{app-eq:min_max_expectation}, a descent-ascent strategy is a natural way to solve this optimization problem. At each iteration, our method runs multiple stochastic gradient ascent steps to estimate the solution of the inner maximization problem. Thanks to this estimation, we can approximate the gradient of the function $h \mapsto W(\hat{\alpha}(h), \hat{\beta})$ that we wish to minimize.
We now go into the detail of our stochastic descent-ascent optimization procedure. 

\paragraph{A new parameterization of the optimization of our problem}

First, in order to avoid projecting $h_n$ on the simplex $\Sigma_K$ at each step of the procedure, we re-parameterized our problem with a soft-max function. Our new optimization problem is now:
\begin{equation}\label{app-eq:min_soft_max}
	\min_{z \in \mathbb{R}^K} W^{\varepsilon}(\Gamma(\sigma(z)), b) = \min_{z \in \mathbb{R}^K}F(z),
\end{equation}
where $F(z)= W^{\varepsilon}(\Gamma(\sigma(z)),b)$, and $\sigma: \mathbb{R}^K \rightarrow \Sigma_K$ is the soft-max function defined as 
\begin{equation}\label{app-eq:soft_max_function}
	\sigma(z)_l = \frac{\exp(z_l)}{\sum_{k=1}^K \exp(z_k)}.
\end{equation}
Then, if we denote $\hat{z}$ a minimizer of \eqref{app-eq:min_soft_max}, we will derive an estimator of $\pi$ as $\hat{\pi} = \sigma(\hat{z})$.

\paragraph{An approximation of the gradient of the objective function $F$}
The gradient of the function $a \mapsto W^{\varepsilon}(a, b)$ is given by:
\begin{equation}
	\frac{\partial}{\partial a}W^{\varepsilon}(a,b) = u^{*}
\end{equation}
where $u^{*}$ is the unique solution to \eqref{app-eq:expectation_wasserstein} centered such that $\sum_{i=1}^I u_i^{*} = 0$. This result is established in Proposition 4.6 of \citep{Peyre}. Applying the chain rule of differentiation to the objective function $F$ of problem \eqref{app-eq:min_soft_max}, we obtain that: 
\begin{equation}\label{app-Gradient_Objective_Function}
	\nabla_{z}W^{\varepsilon}(\Gamma(\sigma(z)), b) = (\Gamma J_{\sigma}(z))^T u_z^*,
\end{equation}
where $\Gamma$ is the linear operator defined in \eqref{app-eq:class_to_observation}, $u_z^*$ denotes a maximizer of \eqref{app-eq:expectation_wasserstein} when the weights of the distribution $\hat{\alpha}$ equal $a=\Gamma \sigma(z)$, and $J_{\sigma}(z)$ is the Jacobian matrix of $\sigma$. For $z \in \mathbb{R}^K$, $J_{\sigma}(z)$ is given by: 
\begin{equation}\label{eq:jacobian_sigma}
	\forall (i,j) \in \{1,...,K\}^2, ~ J_{\sigma}(z)_{i,j} =  \left\{
	\begin{array}{lll}
		\frac{\exp(x_j) \left(\sum_{k=1}^K \exp(x_k) \right) - \exp(2x_j) }{\left(\sum_{k=1}^{K} \exp(x_k)\right)^2}  & \mbox{if } i=j \\
		\\
		-\frac{\exp(x_i + x_j)}{\left( \sum_{k=1}^K \exp(x_k) \right)} & \mbox{otherwise.}
	\end{array}
	\right.
\end{equation}
Using formula \eqref{app-Gradient_Objective_Function}, the gradient of the objective function $F$ at the point $\hat{z}_n$ reads $\nabla_{z}F(\hat{z}_n) = (\Gamma J_{\sigma}(z))^T u_{\hat{z}_n}^*$. We propose to plug an estimate of $u_{\hat{z}_n}^*$ in this last formula where the estimate of $u_{\hat{z}_n}^*$ is computed with the Robbins-Monro \citep{Robbins_Monro} algorithm described in Section \ref{app-sec:stocastic_regularized_wass}. Therefore, the estimate $\widehat{U}_{m+1}^{(n+1)}$ of $u_{\hat{z}_n}^*$ is calculated with the recursive algorithm:
\begin{equation}\label{app-Robbins-Monro step}
	\widehat{U}_{k+1}^{(n+1)} = \widehat{U}_{k}^{(n+1)} + \gamma_{k+1} \nabla_{u}g_{\varepsilon}(Y_{k+1}^{(n+1)}, \widehat{U}_k^{(n+1)}, \hat{z}_n),
\end{equation}
where $\widehat{U}_{0}^{(n+1)}$ is an arbitrary vector in $\mathbb{R}^I$ such that $\langle \widehat{U}_{0}^{(n+1)}, 1_I \rangle = 0$,
$Y_1^{(n+1)},...,Y_{m+1}^{(n+1)}$ are i.i.d random variables sampled from $\hat{\beta}$, and $(\gamma_n)_{n \geq 0}$ is a positive sequence of real numbers decreasing toward zero satisfying condition \eqref{app-eq:step_size_condition}. Finally, our estimate  of $\nabla_{z}W^{\varepsilon}(\Gamma(\sigma(\hat{z}_n)), b)$ is given by:
\begin{equation}\label{app-Estimate gradient}
	\hat{\omega}(\hat{z}_n)= (\Gamma J_{\sigma}(\hat{z}_n))^T \widehat{U}_{m+1}^{(n+1)}.
\end{equation}

\paragraph{An approximated gradient step}
With this stochastic approximation $\hat{\omega}(\hat{z}_n)$ of $\nabla_{z}F(\hat{z}_n)$, we propose an approximated gradient descent to minimize ${F: z \mapsto W^{\varepsilon}(\Gamma \sigma(z), b)}$. Thus, the algorithm considered is given by the recursive procedure 
\begin{equation}\label{app-gradient step}
	\hat{z}_{n+1} = \hat{z}_n - \eta \hat{\omega}(\hat{z}_n).
\end{equation}

Once our minimization procedure is over and we have computed $\hat{z}$ a minimizer of \eqref{app-eq:min_soft_max}, we compute an estimate of the class proportions in the target observations. To do so, we set
\begin{equation}
	\hat{\pi} = \sigma(\hat{z}).
\end{equation}

\begin{algorithm}
	\SetAlgoLined
	
	$z \gets 1_K$\\
	
	\For{$l\leftarrow 1$ \KwTo $n_{out}$}{
		
		$U \gets$ an arbitrary vector in $\mathbb{R}^I$ \\
		
		\For{$k \leftarrow 1$ \KwTo $n_{in}$}{
			
			$Y\sim \hat{\beta}$ \\
			$U \gets U + \gamma_k \nabla_u g_{\varepsilon}(Y, U, \sigma(z))$
			
		}
		
		\tcc{Approximation of the gradient of $z \mapsto  W^{\varepsilon}(\hat{\alpha}(\sigma(z)), \hat{\beta})$}
		$\hat{\omega}(z) \gets (\Gamma J_{\sigma}(z))^{T}U$
		
		$z \gets z - \eta \hat{\omega}(z)$
		
	}
	
	\tcc{Computation of the estimator of the class proportion $\hat{\pi}$}
	\KwRet{$\hat{\pi} = \sigma(z)$}
	
	\caption{Solving $ \min_{z \in \mathbb{R}^K} \max_{u \in \mathbb{R}^I} \mathbb{E}_{Y \sim \hat{\beta}}[g_{\varepsilon, \alpha(h)}(Y,u,\sigma(z))]$}
	
\end{algorithm}

\paragraph{Choice of the parameters for the descent-ascent procedure}\label{app-par:parameters_one}

For this Descent-Ascent procedure, we need to set several parameters. For the step size policy $(\gamma_k)_{k>0}$ of the inner loop, we rely on the recommendation proposed in \citep{Stochastic_Bigot_Bercu}. Therefore, we chose $\gamma_k = \gamma/n^c$ where $c=0.51$ and $\gamma = J\varepsilon/1.9$ with $J$ the cardinal of $\hat{\beta}$. We set the other parameters experimentally. Thus, we chose $n_{out} = 10 000$, $n_{in}=10$, $\eta = 10$ and $\varepsilon = 0.0001$.

\subsection{Minmax swapping procedure}\label{app-subsec:second_method}

\paragraph{An additional regularization term}

In this section we use the ideas of \citep{ballu2020stochastic} to propose an alternative scheme for solving minimization problem \eqref{eq:estimator_proportion}. To this end, we slightly modify problem \eqref{eq:estimator_proportion} by adding the entropic term: 
\begin{equation}\label{app-eq:entropic_two}
	\varphi(h)= \sum_{k=1}^{K} h_k \log(h_k).
\end{equation}
Thus, our new problem is:
\begin{equation}\label{app-eq:estimator_regularization}
	\min_{h \in \Sigma_k} W^{\varepsilon}(\hat{\alpha}(h), \hat{\beta}) + \lambda\varphi(h) = \min_{h \in \Sigma_k} \max_{u \in \mathbb{R}^I} \mathbb{E}[g_{\varepsilon, \alpha}(Y,u,h)] + \lambda \varphi(h),
\end{equation}
where $Y$ is a random variable with distribution $\hat{\beta}$, $g_{\varepsilon, \alpha}$ is defined in \eqref{app-eq:in_the_expectation} and $\lambda$ is a regularization parameter such that $\lambda \geq \varepsilon > 0$. 

\paragraph{Swapping the minimum and the maximum}
By swapping the minimum and the maximum according to Fan's minimax theorem \citep{Fan_theorem} we get: 
\begin{equation}\label{app-eq:swap_minmax}
	\begin{split}
		\min_{h \in \Sigma_k} \max_{u \in \mathbb{R}^I} \mathbb{E}[g_{\varepsilon, \alpha}(Y,u,h)] + \lambda \varphi(h) & =  \max_{u \in \mathbb{R}^I} \min_{h \in \Sigma_k} \mathbb{E}[g_{\varepsilon}(Y,u,h)] + \lambda \varphi(h)\\
		& = \max_{u \in \mathbb{R}^I} \min_{h \in \Sigma_k} \sum_{i=1}^{I} u_i(\Gamma h)_i + A(u) - \varepsilon + \lambda \varphi(h)\\
		& = \max_{u \in \mathbb{R}^I} A(u) + \min_{h \in \Sigma_k} \sum_{i=1}^{I} u_i(\Gamma h)_i + \lambda \varphi(h) - \varepsilon\\
		& = \max_{u \in \mathbb{R}^I} A(u) + J_{\lambda}(u) - \varepsilon,
	\end{split}
\end{equation}
where $A(u) = \varepsilon \sum_{j=1}^J \left( \log(b_j) - \log\left(\sum_{i=1}^I \exp \left( \frac{u_i - c(x_i,y_j)}{\varepsilon} \right) \right) \right)b_j$ and 
\begin{equation}\label{app-eq:inner_problem}
	J_{\lambda}(u) = \min_{h \in \Sigma_k} u_i(\Gamma h)_i + \lambda \varphi(h).
\end{equation}

Arguing e.g.\ as in the proof of  \citep[Proposition 4.1]{Boyer_density_sampling}, the solution $h(u)$ of \eqref{app-eq:inner_problem} is defined by:
\begin{equation}\label{app-eq:inner_solution}
	\forall k \in \{1,...,K\}, ~ (h(u))_k = \frac{\exp\left(-\frac{(\Gamma^T u)_k}{\lambda}\right)}{ \sum_{l=1}^K\exp\left(-\frac{(\Gamma^T u)_l}{\lambda}\right)}.
\end{equation}

By plugging \eqref{app-eq:inner_solution} in \eqref{app-eq:swap_minmax}, problem \eqref{app-eq:estimator_regularization} boils down to the following maximization problem with respect to $u \in \mathbb{R}^d$:
\begin{equation}\label{app-eq:after_swapping}
	\max_{u \in \mathbb{R}^I} \varepsilon \sum_{j=1}^J \left(\log(b_j) - \log \left(\sum_{i=1}^I \exp\left(\frac{u_i-c(x_i,y_j)}{\varepsilon}\right)\right)\right)b_j - \lambda \log\left( \sum_{l=1}^K \exp \left(-\frac{(\Gamma^T u)_l}{\lambda} \right) \right) - \varepsilon.
\end{equation}

\paragraph{An expectation formulation}
This problem can be rewritten as the maximization of an expectation:
\begin{equation}\label{app-eq:alternative_expectation}
	\max_{u \in \mathbb{R}^I}\mathbb{E}[f_{\varepsilon,\lambda, \alpha}(Y,u)],
\end{equation}
where $Y$ is a random variable with distribution $\hat{\beta}$, and for $y_j \in \mathbb{R}^d$ an observation of $Y$, and $u \in \mathbb{R}^I$,
\begin{equation}\label{app-eq:expectation_two}
	f_{\varepsilon, \lambda, \alpha}(y_j, u) = \varepsilon  \left(\log(b_j) - \log \left(\sum_{i=1}^I \exp\left(\frac{u_i-c(x_i,y_j)}{\varepsilon}\right)\right)\right) - \lambda \log\left( \sum_{l=1}^K \exp \left(-\frac{(\Gamma^T u)_l}{\lambda} \right) \right) - \varepsilon.
\end{equation}
The fact that for all $y \in \mathbb{R}^d$ the function $f_{\varepsilon, \lambda}(y,.)$ is concave and the expectation form \eqref{app-eq:alternative_expectation} of problem \eqref{app-eq:after_swapping} lead us to estimate the optimal vector $u^*$ by the Robbins-Monro algorithm given, for all $n \geq 0$ by 
\begin{equation}\label{app-eq:robbin_monro_two}
	\widehat{U}_{n+1} = \widehat{U}_n + \gamma_{n+1}\nabla_{u}f_{\varepsilon, \lambda, \alpha}(Y_{n+1},\widehat{U}_n),
\end{equation}
where the initial value $\hat{U}_{0}$ is a random vector which can be arbitrarily chosen, $Y_1,...,Y_{n+1}$ are i.i.d random variables sampled from $\hat{\beta}$ ,and $(\gamma_n)_{n \geq 0}$ is a positive sequence of real numbers decreasing toward zero satisfying condition \eqref{app-eq:step_size_condition}. Moreover, the gradient of $f_{\varepsilon, \lambda}$ can be easily calculated. Indeed, for $y_j \in \mathbb{R}^d$ an observation of $Y$, and $u \in \mathbb{R}^I$, $\nabla_{u}f_{\varepsilon, \lambda}(y_j,u)$ is defined by $\forall i_0 \in \{1,...,I\}$,
\begin{equation}\label{app-eq:gradient_alternative}
	(\nabla_{u}f_{\varepsilon, \lambda}(y_j,u))_{i_0} =  \frac{\sum_{l=1}^K \Gamma_{i_0,l} \exp \left(-\frac{(\Gamma^T u)_l}{\lambda}\right)}{\sum_{k=1}^K \exp \left(-\frac{(\Gamma^T u)_k}{\lambda}\right)
	}
	- \frac{\exp\left(\frac{u_{i_0}-c(x_{i_0},y_j)}{\varepsilon}\right)}{\sum_{i=1}^I \exp\left(\frac{u_i-c(x_i,y_j)}{\varepsilon}\right)}.
\end{equation}
Once the algorithm has converged, and we get $\hat{U}$, a satisfactory approximation of a maximizer of problem \ref{app-eq:alternative_expectation}, one can compute an estimate of the class proportions by setting:
\begin{equation}\label{app-eq:second_estimator}
	\hat{\pi} = h(\hat{U}),
\end{equation}
where $h(u)$ is defined in \ref{app-eq:inner_solution}.

\begin{algorithm}[!ht]
	\SetAlgoLined
	
	$U \gets 0_I$ \\
	
	\For{$l\leftarrow 1$ \KwTo $n$}{
		
		$Y\sim \hat{\beta}$ \\
		
		$U \gets U + \gamma_l \nabla_u f_{\varepsilon, \lambda}(Y, U)$
	}
	
	\tcc{Computation of the estimator of the class proportion $\hat{p}$}
	
	$ \hat{p} \gets h(U)$\\
	
	\KwRet{$\hat{p}$}
	
	\caption{Solving $ \max_{u \in \mathbb{R}^I} \mathbb{E}_{Y \sim \hat{\beta}}[f_{\varepsilon, \lambda}(Y,u)]$}
	
\end{algorithm}

\paragraph{Choice of the parameters for the min-max swapping procedure}

For this second procedure we have to set several parameters.  To satisfy condition \eqref{app-eq:step_size_condition} for the step size policy $(\gamma_n)_{n > 0}$ of the Robbins-Monro algorithm, we chose $\gamma_n = \gamma/n^c$ where $\gamma=5$ and $c=0.99$. The other parameters have been set empirically. We chose $\lambda=0.0001$ and $\varepsilon=0.0001$. The number of iterations for this stochastic ascent algorithm has been set to $n=10000$.

\subsection{Preprocessing of the cytometry data}

Finally, note that when applying our method to estimate the class proportions on real flow cytometry data, a simple pre-processing is required. First, the signal processing of the cytometer can induce some contrived negative values of light intensity. To undo this effect, we merely threshold those few negative values at zero. Second, to settle the parameters of our algorithms, in particular $\varepsilon$, we need to bound the displacement cost. To do so, we scale the data such that: $\forall i \in \{1,...,I\},~ X_i^s \in [0,1]^d$ and $\forall j \in \{1,...,J\}, ~ X_j^t \in [0,1]^d$.

\subsection{Comparison of the minimization procedures}

\begin{figure}[!ht]
	\centering
	\includegraphics[width=10cm]{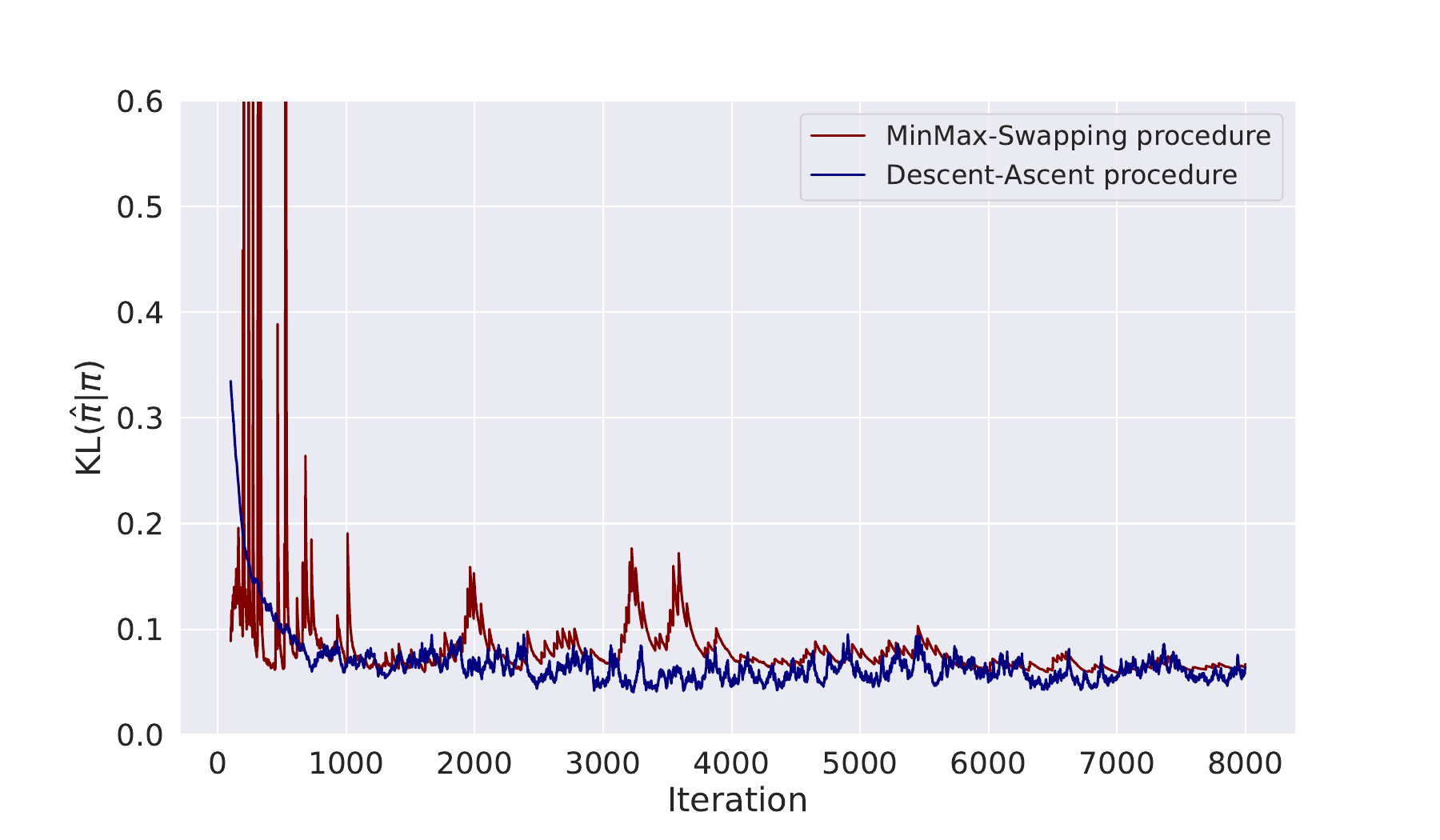}
	\caption{\footnotesize\textbf{Evolution of the Kullback-Leibler divergence between the estimated proportions $\hat{\pi}$ and the manual gating benchmark $\pi$.}}
	\label{fig:minimization_procedure} 
\end{figure}

As said in Subsection \ref{subsec:kullback_divergence}, we assess the performance of our estimator $\hat{\pi}$ by computing the Kullback-Leibler divergence between $\hat{\pi}$ and $\pi$ the benchmark class proportions.
Figure \ref{fig:minimization_procedure} displays the evolution of the Kullback-Leibler divergence along the iterations of the two minimization procedures. For the descent-ascent procedure, the iterations displayed correspond to the iterations of the outer loop. This comparison has been realized when the source data set is Stanford1A segmented into 10 classes. The target data set is Stanford3A where we try to estimate the proportions of the 10 cell sub-populations.

Both algorithms have been implemented in the Python package available at \url{https://github.com/sistm/CytOpt-python}. From our numerical experiments, we have found that the sequence of estimates $\hat{\pi}$ produced by the second algorithm described in Section \ref{app-subsec:second_method} levels off approximately ten times faster than the sequence of estimates produced with the descent-ascent procedure described in Section \ref{app-subsec:first_method}. This computational time gain can be accounted by the simple loop complexity of this second algorithm. As in practice both procedures seem to provide close estimates, all the results reported where produced with the second strategy, referred to as the "min-max swapping" strategy.

\section{Additional figures}

\begin{figure}[!ht]
	\centering
	\includegraphics[width=10cm]{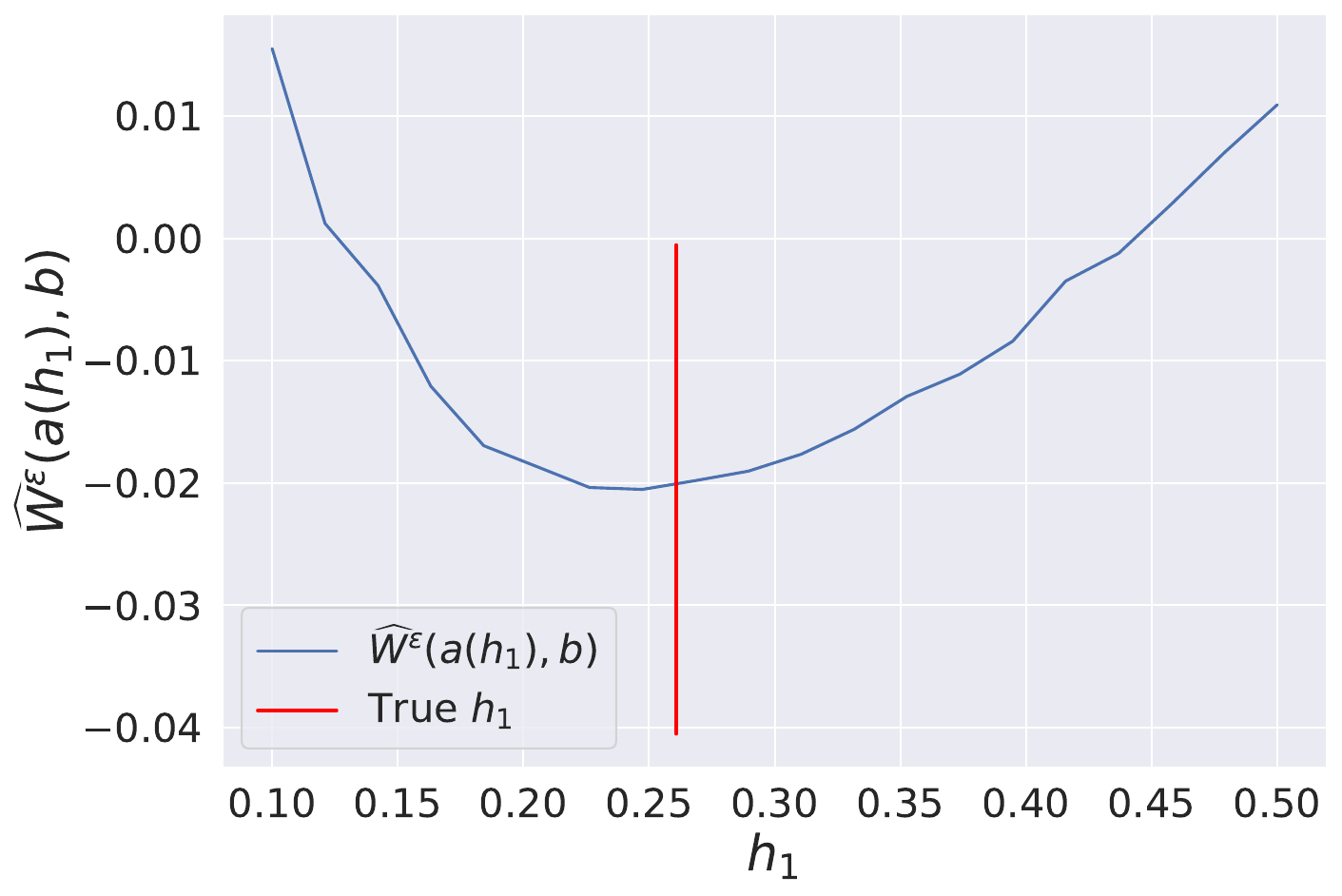}
	\caption{\footnotesize\textbf{Approximation of the function $h_1 \mapsto W^{\varepsilon}(\hat{\alpha}(h), \hat{\beta})$, for $h = (h_1, 1-h_1)$, $h_1$ represents the weight associated to the CD8 cells.} The approximation of $W^{\varepsilon}(\hat{\alpha}(h), \hat{\beta})$ is produced using the estimator \eqref{app-eq:estimate_wasserstein}.}
	\label{app-fig:wasserstein_evolution} 
\end{figure}

To assess the relevance of our method, we present in Figure \ref{app-fig:wasserstein_evolution} the evolution of the regularized Wasserstein distance as a function of the weights associated to each class in the source distribution. To this end, we evaluate the function $F: h_1 \mapsto W^{\varepsilon}(\hat{\alpha}(h), \hat{\beta})$, where $h=(h_1, 1-h_1)$, on a finite grid $\mathcal{H} = \{ h^{(1)},...,h^{(m)} \}$. For $h_1 \in \mathcal{H}$ we approximate $W^{\varepsilon}(\hat{\alpha}(h), \hat{\beta})$ by the estimator $\widehat{W}_n$ defined in equation \eqref{app-eq:after_swapping}. It can be observed that the regularized Wasserstein distance decreases as the class proportions of the source data set get closer to the class proportions of the target data set.

\clearpage
\bibliography{Biblio_First_Article}

\end{document}